\PassOptionsToPackage{table,xcdraw}{xcolor}
\documentclass[a4paper,11pt]{article}

\pdfoutput=1 

\usepackage[shortlabels]{enumitem}
\usepackage{pst-all}	
\usepackage{physics}
\usepackage{blkarray}
\usepackage{mathtools}
\usepackage{booktabs}
\usepackage{jcappub} 
\usepackage[T1]{fontenc} 
\usepackage{cleveref}
\usepackage{tabularx} 
\usepackage[normalem]{ulem}
\usepackage{xcolor}
\newif\ifshowchanges
\showchangestrue
\ifshowchanges
  \newcommand{\old}[1]{{\color{red}\sout{#1}}}
  \newcommand{\new}[1]{{\color{blue}#1}}
  \newcommand{\flag}[1]{{\color{magenta}\textbf{[#1]}}}
\else
  \newcommand{\old}[1]{}
  \newcommand{\new}[1]{#1}
  \newcommand{\flag}[1]{}
\fi
\usepackage{multirow}
\usepackage{tikz}
\usetikzlibrary{matrix, fit, positioning,arrows.meta,arrows, ,decorations.pathreplacing,shapes.geometric,backgrounds}

\title{Cosmological constraints from the DESI DR1 Bispectrum Full-Shape and DR2 BAO}

\usepackage{orcidlink}

\affiliation{Affiliations are in Appendix \ref{sec:affiliations}}
\emailAdd{daniel.forerosanchez@icc.ub.edu}

\author[a]{{D.~Forero-Sánchez}\orcidlink{0000-0001-5957-332X},}
\author[a,av]{{S.~Novell-Masot},}
\author[b,c,a]{{H.~Gil-Mar\'in}\orcidlink{0000-0003-0265-6217},}
\author[d,a]{{L.~Verde}\orcidlink{0000-0003-2601-8770},}
\author[e]{{J.~Aguilar},}
\author[f]{{S.~Ahlen}\orcidlink{0000-0001-6098-7247},}
\author[g,h]{{D.~Bianchi}\orcidlink{0000-0001-9712-0006},}
\author[e]{{A.~Brodzeller}\orcidlink{0000-0002-8934-0954},}
\author[i]{{D.~Brooks},}
\author[c,j]{{F.~J.~Castander}\orcidlink{0000-0001-7316-4573},}
\author[k]{{S.~Cole}\orcidlink{0000-0002-5954-7903},}
\author[l]{{A.~de la Macorra}\orcidlink{0000-0002-1769-1640},}
\author[m]{{J.~Della~Costa}\orcidlink{0000-0003-0928-2000},}
\author[n,o]{{Biprateep~Dey}\orcidlink{0000-0002-5665-7912},}
\author[i]{{P.~Doel},}
\author[e,p]{{S.~Ferraro}\orcidlink{0000-0003-4992-7854},}
\author[d,q]{{A.~Font-Ribera}\orcidlink{0000-0002-3033-7312},}
\author[r,s]{{J.~E.~Forero-Romero}\orcidlink{0000-0002-2890-3725},}
\author[t]{{Satya~{Gontcho A Gontcho}}\orcidlink{0000-0003-3142-233X},}
\author[u]{{G.~Gutierrez},}
\author[v]{{C.~Hahn}\orcidlink{0000-0003-1197-0902},}
\author[w,x]{{H.~K.~Herrera-Alcantar}\orcidlink{0000-0002-9136-9609},}
\author[y,z,aa]{{K.~Honscheid}\orcidlink{0000-0002-6550-2023},}
\author[ab,ac]{{D.~Huterer}\orcidlink{0000-0001-6558-0112},}
\author[ad]{{M.~Ishak}\orcidlink{0000-0002-6024-466X},}
\author[ae]{{D.~Kirkby}\orcidlink{0000-0002-8828-5463},}
\author[e]{{A.~Kremin}\orcidlink{0000-0001-6356-7424},}
\author[i]{{O.~Lahav}\orcidlink{0000-0002-1134-9035},}
\author[aa]{{C.~Lamman}\orcidlink{0000-0002-6731-9329},}
\author[e]{{M.~Landriau}\orcidlink{0000-0003-1838-8528},}
\author[af]{{L.~Le~Guillou}\orcidlink{0000-0001-7178-8868},}
\author[e]{{M.~E.~Levi}\orcidlink{0000-0003-1887-1018},}
\author[ag,q]{{M.~Manera}\orcidlink{0000-0003-4962-8934},}
\author[m]{{A.~Meisner}\orcidlink{0000-0002-1125-7384},}
\author[d,q]{{R.~Miquel},}
\author[ah]{{J.~Moustakas}\orcidlink{0000-0002-2733-4559},}
\author[ai]{{S.~Nadathur}\orcidlink{0000-0001-9070-3102},}
\author[o]{{J.~ A.~Newman}\orcidlink{0000-0001-8684-2222},}
\author[aj,ak]{{G.~Niz}\orcidlink{0000-0002-1544-8946},}
\author[x,e]{{N.~Palanque-Delabrouille}\orcidlink{0000-0003-3188-784X},}
\author[al,am,an]{{W.~J.~Percival}\orcidlink{0000-0002-0644-5727},}
\author[ao]{{F.~Prada}\orcidlink{0000-0001-7145-8674},}
\author[ap]{{I.~P\'erez-R\`afols}\orcidlink{0000-0001-6979-0125},}
\author[aq]{{G.~Rossi},}
\author[ar,as,at]{{L.~Samushia}\orcidlink{0000-0002-1609-5687},}
\author[au]{{E.~Sanchez}\orcidlink{0000-0002-9646-8198},}
\author[e]{{D.~Schlegel},}
\author[ab,ac]{{M.~Schubnell},}
\author[e]{{J.~Silber}\orcidlink{0000-0002-3461-0320},}
\author[ac]{{G.~Tarl\'{e}}\orcidlink{0000-0003-1704-0781},}
\author[m]{{B.~A.~Weaver},}

\newcommand{\Mpch}{\,h^{-1}\,{\rm Mpc}}
\newcommand{\hMpc}{\,h\,{\rm Mpc}^{-1}}

\newcommand{\lcdm}{$\Lambda$CDM}
\newcommand{\wowacdm}{$w_0w_a$CDM}
\newcommand{\nulcdm}{$\nu\Lambda$CDM}
\newcommand{\nuwowacdm}{$\nu w_0w_a$CDM}

\newcommand{\lya}{Ly$\alpha$}

\abstract{We present cosmological constraints from the combination of DESI DR1 full‑shape measurements, including the LRG bispectrum, and DESI DR2 BAO data. The joint analysis accounts for cross‑covariance using mocks, while ShapeFit compression mitigates prior volume effects that hinder beyond-$\Lambda$CDM analyses. In $\Lambda$CDM, the DESI1.5 (P+B) analysis yields  $\sigma_8=0.868\pm0.026$ and $S_8=0.880\pm0.028$, compared with $\sigma_8=0.822\pm0.034$ and $S_8=0.827\pm0.036$ in the previous power-spectrum-only DESI1.5 analysis \cite{desi1.5}. The corresponding marginalized uncertainties are reduced by 24\% and 22\%, respectively. The improved sensitivity to the clustering amplitude is consistent with the impact of the bispectrum demonstrated in the matched DR1 (P) versus (P+B) analysis of Ref.~\cite{novell-masot26}. For $w_0w_a$CDM, DESI‑only analyses with the bispectrum shift dark energy parameters toward $\Lambda$CDM, staying consistent with a cosmological constant within $1\sigma$ in the two-dimensional $w_0$--$w_a$ plane. Adding CMB creates a preference for evolving dark energy: DESI+CMB (P+B) shows a $2.8\sigma$ deviation from $\Lambda$CDM. Including DES‑Dovekie supernovae alone reduces this to $1.6\sigma$, while the full combination DESI+CMB+DES‑Dovekie gives $3.1\sigma$, driven primarily by the CMB. The bispectrum consistently weakens evidence for time‑varying dark energy relative to power‑spectrum–only analyses. The bispectrum also enhances sensitivity to massive neutrinos: in \new{the} DESI‑only analysis, the power‑spectrum–only posterior for $\sum m_\nu$ is consistent with zero, whereas adding the bispectrum yields a mean of  $0.26\pm0.18$~eV and a $95\%$ upper limit of  $0.59$~eV, shifting the peak into the positive region and agreeing with oscillation lower bounds. For modified gravity, the bispectrum further constrains  $\mu_0 = 0.12\pm0.51$ from DESI‑only data, consistent with general relativity. Our analysis shows that accounting for cross‑dataset covariances and avoiding prior volume effects yields robust constraints, with the bispectrum raising amplitude parameters and tightening their uncertainties.}

\crefname{figure}{figure}{figures}
\Crefname{figure}{Figure}{Figures}
\Crefname{table}{Table}{Tables}

\begin{document}
\maketitle
\flushbottom
\ifshowchanges
\fi

\section{Introduction}
\label{sec:intro}

The Dark Energy Spectroscopic Instrument (DESI) \cite{levi2013desi, Levi2019,DESI2016a.Science} has surveyed the sky with unprecedented precision, mapping the large-scale structure of the Universe from redshifts $z\sim0.1$ to $z\sim4$. Located at the 4-meter Mayall Telescope, DESI is a highly multiplexed, fiber-fed spectrograph capable of simultaneously observing $\sim$5000 targets across a 3$^\circ$ field of view \cite{DESI2016b.Instr, Corrector.Miller.2023, Spectro.Pipeline.Guy.2023,FiberSystem.Poppett.2024,DESI2022.KP1.Instr,SurveyOps.Schlafly.2023}. The survey targets four primary tracers: Bright Galaxies (BGS) \citep{BGSPrelim.RuizMacias.2020,BGS.TS.Hahn.2023}, Luminous Red Galaxies (LRG) \citep{LRG.TS.Zhou.2023}, Emission Line Galaxies (ELG) \citep{ELGPrelim.Raichoor.2020,ELG.TS.Raichoor.2023}, and quasars (QSO) \citep{QSOPrelim.Yeche.2020,QSO.TS.Chaussidon.2023}, the latter also enabling Lyman-$\alpha$ forest measurements. Over its extended 8-year timeline, DESI will obtain spectra for over 60 million objects, with the First Data Release (DR1) now publicly available \cite{DESI2024.I.DR1} and Second Data Release (DR2) \cite{DESI.DR2.DR2} expected in Spring 2027.

A cornerstone of the DESI science program is mapping the Baryon Acoustic Oscillation (BAO) feature in the clustering of each tracer. This provides robust constraints on the cosmic expansion history, particularly on $\Omega_m$ and the combination $r_{\rm d}H_0$ within $\Lambda$CDM. Recent BAO measurements from DR1 and DR2, when combined with Cosmic Microwave Background (CMB) and Supernovae (SNe) data, have hinted at a possible time-evolving dark energy component \citep{DESI2024.VI.KP7A,DESI.DR2.BAO.cosmo,DESI.DR2.BAO.lya}, though recent reanalyses of the SNe samples \cite{DESreanalysis} have decreased the discrepancies between \lcdm{} and the data.

While BAO is a powerful probe of background evolution, additional cosmological information resides in the Full-Shape (FS) of the galaxy power spectrum, which is sensitive to structure growth, neutrino masses, and modified gravity. The initial DESI FS analyses \citep{DESI2024.V.KP5,DESI2024.VII.KP7B,KP7s1-MG} employed a ``full modeling'' (FM) approach, directly fitting perturbation-theory predictions to the two-point statistics. Although FM exploits the broadband shape to tighten constraints, it often requires informative priors on nuisance parameters to obtain reliable constraints in extended models. In DESI-only analyses of models beyond $\Lambda$CDM, the cosmological interpretation of  FM constraints can become unreliable due to significant prior volume effects, particularly for dynamical dark energy parameterizations. 

Our previous work \cite{desi1.5} combined DESI DR1 power-spectrum full-shape measurements with DR2 BAO data using the ShapeFit compression methodology \citep{Brieden21}. ShapeFit extracts scale dilation parameters ($q_\parallel$ and $q_\perp$), the logarithmic growth rate $f$, the amplitude of the power spectrum $\sigma_8$, and shape parameters ($m$, $n$) that capture slope variations around the matter-radiation equality scale. This approach serves as a robust middle ground between standard BAO-RSD compression and full modeling, decoupling data compression from cosmological inference and enabling reliable reinterpretation across various models without the computational overhead or prior sensitivity of FM. In particular, ShapeFit circumvents the prior-volume effects that have hindered previous full-modeling analyses of extended cosmologies, yielding the first DESI+BBN+$n_s$ constraints on models where FM has previously failed.

The high number density of DESI tracers provides the statistical power necessary to probe higher-order correlations. The bispectrum, in particular, has long been recognized as a key statistic to break parameter degeneracies, tighten cosmological constraints, and provide stringent tests of \lcdm{}. 

This work is part of a series of papers. Ref.~\cite{novell-masot25} validated our SF pipeline for DESI DR1 power spectrum and bispectrum measurements, thoroughly testing various cosmology variations, halo occupation distribution models and observational systematics, thus deriving a systematic error budget following DESI's unblinding policies. Furthermore, it tests for variations in the analyses, such as galaxy power spectrum modeling, scale cuts and multipole choices, finding results consistent with DESI's fiducial choices in \cite{DESI2024.V.KP5}.

Ref.~\cite{novell-masot26} builds on these results and presents the cosmological interpretation of the DR1 SF measurements of \cite{novell-masot25} both with and without the addition of the bispectrum. We find that these are consistent with the DESI collaboration's original findings while delivering tighter constraints and successfully disentangling the $f$ and $\sigma_8$ parameters. Moreover, comparing the power-only and power-plus-bispectrum measurements from \cite{novell-masot25} we observe that the bispectrum information is largely responsible for the improvements in the constraining power and shifts in the dark-energy parameter space.

On the other hand, \cite{desi1.5} introduces the methodology to combine DR1 full-shape and DR2 BAO measurements in a robust way, taking into account DR1$\times$DR2 correlations using mocks, also demonstrating that the use of SF compression largely mitigates projection effects in extended cosmological models. 

In this work, we update our recent results on DESI FS \cite{desi1.5} using the bispectrum information derived previously in \cite{novell-masot25}. Here, we combine the DESI DR1 LRG bispectrum with DR2 BAO, while, for the first time, properly accounting for their cross-covariance using mock-based estimates. Our aim is therefore to provide the most comprehensive DESI-only clustering constraints across a variety of cosmological models.

This work is organized as follows. \Cref{sec:data} introduces the data used, both DESI and external, then \cref{sec:methods} reviews the methods used to combine DR1 and DR2 data while accounting for their correlation as well as revisiting the BAO and ShapeFit analyses. \Cref{sec:results} shows and discusses the results in terms of sample combinations and cosmology. Finally, we conclude in \cref{sec:conclusions}.

\section{Data and Mocks}
\label{sec:data}

\definecolor{lightgray}{HTML}{EFEFEF} 

\begin{table}[h]
\centering
\scalebox{0.9}{\begin{tabular}{lllcccc}
\toprule
Tracer & Redshift Range & Data Release & $N_{\text{tracer}}$ & $z_{\text{eff}}$ & $V_{\text{eff}}$ (Gpc$^3$) & Use \\
\midrule

\multirow{2}{*}{BGS} & \multirow{2}{*}{0.1 – 0.4} & \cellcolor{lightgray}DESI DR1 & \cellcolor{lightgray}300,017 & \cellcolor{lightgray}0.295 & \cellcolor{lightgray}1.7 & \cellcolor{lightgray} --- \\
                      &                             & DESI DR2 & 1,188,526 & 0.295 & 3.8 & BAO\\ 
\multirow{2}{*}{LRG1} & \multirow{2}{*}{0.4 – 0.6} & \cellcolor{lightgray}DESI DR1 &\cellcolor{lightgray} 506,905 & \cellcolor{lightgray}0.510 &\cellcolor{lightgray} 2.6 & \cellcolor{lightgray} (P+B)-FS \\
                      &                             & DESI DR2 & 1,052,151 & 0.510 & 4.9 & BAO\\ 
\multirow{2}{*}{LRG2} & \multirow{2}{*}{0.6 – 0.8} & \cellcolor{lightgray}DESI DR1 &\cellcolor{lightgray} 771,875 & \cellcolor{lightgray}0.706 &\cellcolor{lightgray} 4.0 & \cellcolor{lightgray} (P+B)-FS \\
                      &                             & DESI DR2 & 1,613,562 & 0.706 & 7.6 & BAO\\ 
\multirow{2}{*}{LRG3} & \multirow{2}{*}{0.8 – 1.1} & \cellcolor{lightgray}DESI DR1 & \cellcolor{lightgray}859,824 & \cellcolor{lightgray}0.920 & \cellcolor{lightgray}5.0 & \cellcolor{lightgray}(P+B)-FS\\
                      &                             & DESI DR2 & 1,802,770 & 0.920 & 9.8 & BAO\\
\multirow{2}{*}{ELG2} & \multirow{2}{*}{1.1 – 1.6} & \cellcolor{lightgray}DESI DR1 & \cellcolor{lightgray}1,415,687 &\cellcolor{lightgray} 1.317 &\cellcolor{lightgray} 2.7  & \cellcolor{lightgray} P-FS\\
                      &                             & DESI DR2 & 3,797,271 & 1.317 & 8.3 & BAO\\ 
\multirow{2}{*}{QSO} & \multirow{2}{*}{0.8 – 2.1} & \cellcolor{lightgray}DESI DR1 &\cellcolor{lightgray} 856,652 &\cellcolor{lightgray} 1.491 &\cellcolor{lightgray} 1.5 & \cellcolor{lightgray}P-FS\\
                      &                             & DESI DR2 & 1,461,588 & 1.491 & 2.7 & BAO\\ 
\multirow{2}{*}{Ly$\alpha$} & \multirow{2}{*}{1.77 – 4.16} & \cellcolor{lightgray}DESI DR1 &\cellcolor{lightgray} 709,565 & \cellcolor{lightgray}2.330 &\cellcolor{lightgray} --- & \cellcolor{lightgray} FS (AP) \\
                            &                             & DESI DR2 & 1,289,874 & 2.330 & --- & BAO\\
\bottomrule
\end{tabular}}
\caption{Statistics for each of the DESI tracer types used
for the DESI DR1 FS and DR2 BAO measurements presented in this paper. The last column shows whether the sample is used in this analysis and in what form, either BAO or Full-Shape (FS) as well as the statistics used, either the power spectrum only (P) or both power spectrum and bispectrum (PB). The FS information from \lya{} sample from DR1 is used to provide constraints on the Alcock-Paczynski (AP) effect rather than to constrain growth as is the case for the other tracers.}
\label{tab:desisamples}
\end{table}

\subsection{DESI dataset}
The cosmological analyses presented here rely on two successive public data releases from the Dark Energy Spectroscopic Instrument (DESI). Data Release 1 (DR1), made public after the first year of operations (concluded June 2022), encompasses over 6 million objects across 7500 square degrees, representing roughly 44\% of the final survey footprint \cite{DESI2024.I.DR1}. This already surpassed the tracer count of the Sloan Digital Sky Survey III Baryon Oscillation Spectroscopic Survey  (BOSS)  by more than a factor of two \cite{BOSS,eBOSS1}. Building on DR1, Data Release 2 (DR2), incorporates three years of DESI observations, completed in April 2024. DR2 expands the sky coverage by approximately 60\% relative to DR1 and more than doubles the number of usable objects to over 14 million \cite{KP3s15-Ross}. 

This rapid accumulation of high-quality spectra is enabled by DESI's highly multiplexed optical design. The instrument utilizes approximately 5000 optical fibers that are robotically positioned on the focal plane, allowing for the simultaneous acquisition of thousands of spectra. This architecture delivers a substantial increase in observational efficiency compared to previous spectroscopic campaigns \cite{FiberSystem.Poppett.2024,FocalPlane.Silber.2023}.

DESI's observing strategy divides the survey into two complementary programs tailored to different sky brightness conditions. The Bright program operates during nights with significant lunar illumination to target nearby galaxies comprising the Bright Galaxy Survey (BGS) \cite{BGS.TS.Hahn.2023}. Conversely, the Dark program takes advantage of darker skies to observe more distant tracers, including Luminous Red Galaxies (LRG), Emission Line Galaxies (ELG), and quasars (QSO) \cite{LRG.TS.Zhou.2023,ELG.TS.Raichoor.2023,QSO.TS.Chaussidon.2023}. 

Over its planned eight-year runtime, which commenced in May 2021, DESI will ultimately map roughly 17,000 square degrees, probing the large-scale structure across a wide redshift interval from $z\simeq0.1$ to $z\simeq4.2$. By combining these diverse tracer populations, the survey samples the cosmic expansion history at multiple epochs. Detailed metrics regarding the redshift coverage, effective redshifts, survey volumes, and object counts for each sample are compiled in \Cref{tab:desisamples}.


For the purposes of our analysis, the distinct redshift bins within these releases are treated as statistically independent and uncorrelated.

Throughout this work we combine the DESI DR1 Full-Shape and DESI DR2 BAO data. This combination we call DESI1.5 (following the nomenclature introduced in \cite{desi1.5}). When using DR2 BAO data alone, this is labeled as BAO2. 

\subsection{EZmocks}

To quantify the cross-covariance between the DR1 and DR2 datasets, we employ the DESI EZmock suite \cite{KP3s8-Zhao,Chuang2015EZmock,Zhao2021}, which comprises one thousand mocks per galactic cap for each of the DESI dark time samples. DR1 is a subset of DR2, implying substantial overlap in the underlying spectroscopic targets. Nevertheless, in \cite{desi1.5} we show that for power spectrum-only FS analyses, the cross-correlation has negligible effects in the cosmology constraints. This slightly counterintuitive finding is likely due to the DR2 BAO measurements   having dominant statistical weight and the dimensionality reduction inherent to the compressed ShapeFit parameterization: both effects suppress the cross-correlations.

However, the DR1-DR2 cross correlation  may not be negligible when including  higher-order statistics. Thus, here we compute the full cross-covariance by extracting joint power spectrum and bispectrum measurements from the mock catalogues and cross-correlating them with the DR2 BAO results in the compressed parameter space. A comprehensive description of the covariance construction pipeline is provided in our preceding analysis \cite{desi1.5} and is briefly summarized in \cref{sec:methods_combination}, while detailed specifications regarding the mock generation and validation procedures are documented in \cite{KP3s8-Zhao,Zhao2021,Chuang2015EZmock,KP4s6-Forero-Sanchez}.

\subsection{External datasets and priors}\label{sec:external}

Our external data selection closely mirrors the prescriptions adopted in the official DESI DR2 BAO \citep{DESI.DR2.BAO.cosmo}, DR1 Full-Shape \citep{DESI2024.VII.KP7B} and previous relevant analyses \cite{novell-masot25,novell-masot26,desi1.5}. We focus however on analyses that predominantly use DESI information in order to highlight its specific role. Our baseline analyses are therefore done imposing minimal external priors on the physical baryon density $\omega_{\rm b}$ and the primordial scalar spectral index $n_s$.

The $\omega_{\rm b}$ prior is informed by Big Bang Nucleosynthesis (BBN) constraints from \cite{Schonenberg2024_BBN}, which combine theoretical predictions of light-element abundances with updated uncertainties in relevant nuclear cross-sections. Following the standard DESI practice, we utilize the joint $\omega_{\rm b}$--$N_{\rm eff}$ measurement from \cite{Schonenberg2024_BBN}, fixing the effective number of relativistic species to $N_{\rm eff}=3.044$. This yields the following bivariate Gaussian prior:\footnote{Hereafter, $\mathcal{N}(\mu, \Sigma^2)$ denotes a normal distribution with mean $\mu$ and (co)variance $\Sigma^2$.}
\begin{equation}
    \mathcal{L}(\omega_{\rm b}) = \mathcal{N}\left(\begin{pmatrix}0.02196 \\ 2.944\end{pmatrix}, 
        \begin{pmatrix}4.03\times 10^{-7} & 7.30\times 10^{-5} \\ 7.30\times 10^{-5} & 4.53\times10^{-2}\end{pmatrix}\right).
\end{equation}

For the spectral index $n_s$, we employ a Gaussian prior centered on the Planck (2018) measurement \citep{planck_collaboration_planck_2018}. To ensure a conservative baseline, we deliberately inflate the associated uncertainty by a factor of ten (denoted $n_{s10}$):
\begin{equation}
    \mathcal{L}_{n_{s10}}(n_s) = \mathcal{N}(0.9649, 0.042^2).
\end{equation}

When CMB data are included, we utilize the Planck (2018) PR3 temperature and polarization power spectra (TT, TE, and EE) \citep{planck_collaboration_planck_2018}. Specifically, we adopt the \texttt{SimAll Commander} likelihood for multipoles $\ell < 30$ and the \texttt{Plik} likelihood for $\ell \geq 30$. For extended cosmological scenarios involving massive neutrinos or non-zero spatial curvature, we switch to the \texttt{CamSpec} likelihoods from Planck PR4 \citep{PlanckPR41,PlanckPR42}, consistent with the DR2 BAO analysis protocol. In all CMB-inclusive configurations, we supplement the primary temperature/polarization likelihood with reconstructed CMB lensing information. The lensing potential is extracted from the connected four-point function of the CMB maps, combining the NPIPE-based \textit{Planck} PR4 reconstruction \cite{plancklens} with the Atacama Cosmology Telescope Data Release 6 (ACT-DR6) \citep{act1,act2,act3}. Throughout the remainder of this work, the combination of Planck TT/TE/EE and Planck+ACT lensing data is collectively referred to as ``CMB''.

Additionally, we make use of the updated Type Ia supernova (SNe Ia) sample from the Dark Energy Survey, referred to as DES‑Dovekie, presented by Ref.~\cite{DESreanalysis}. This dataset is a reanalysis of the DES 5‑year SN sample (DES‑SN5YR; \cite{desy5,desy52}), incorporating several important improvements: an enhanced photometric cross‑calibration using new white dwarf observations \cite{Boyd25} and Gaia spectroscopy, a retrained SALT3 light‑curve model (SALT3.DOV), and a correction to a long‑standing approximation in the Fitzpatrick dust colour law \cite{Fitzpatrick99}. The sample comprises approximately 1600 likely SNe Ia from DES (photometrically classified) and about 200 low‑redshift SNe from external surveys (CfA3, CfA4, CSP, Foundation), all with spectroscopic host‑galaxy redshifts.

DES‑Dovekie provides distance moduli and a full systematic covariance matrix, and has been shown to yield cosmological constraints that are fully consistent with Pantheon+ and Union3 while reducing the previously reported preference for evolving dark energy to a weak level \cite{DESreanalysis}. In our analysis, we use the DES‑Dovekie SNe in combination  with CMB and BAO data to obtain robust constraints on extended cosmological models.

\section{Methods}
\label{sec:methods}

Our data combination procedure is based on computing a DR1 FS $\times$ DR2 BAO sample covariance estimated from EZmocks in the compressed parameter space. This requires performing the BAO analysis on a set of DR2 EZmocks and the FS analysis (including the bispectrum) on the phase-matched DR1 EZmocks. Relative to computing cross covariances in the space of $k$-bandpowers, working in this compressed parameter space allows us to use orders of magnitude fewer mocks. This point is especially relevant when using the bispectrum, where the number of triangle configurations increases drastically relative to the 2-point only analyses.
We summarize the BAO, Shapefit  and covariances pipelines below, highlighting the differences in the FS modeling due to the inclusion of the bispectrum and refer the reader to \cite{GilMarin2022_comb,novell-masot25,desi1.5} for in-depth explanations.

\subsection{BAO Modeling}
\label{sec:bao}

The BAO are described by a pair of compressed parameters that quantify shifts in the BAO peak position relative to a fiducial cosmology template. The line-of-sight and transverse dilation factors are defined as\footnote{The definitions here omit the ``BAO'' superscript for brevity; the same forms hold for the SF compressed parameters. We use this $q$-notation for consistency with other DESI papers, though these are often referred to as $\alpha$ in the literature.} 
\begin{align}
    q_\parallel(z) &= \frac{H^{\rm fid}(z)}{H(z)}\frac{r_d^{\rm fid}}{r_d}, &
    q_\perp(z) &= \frac{D_M(z)}{D_M^{\rm fid}(z)}\frac{r_d^{\rm fid}}{r_d},
\end{align}
where $H(z)$ is the Hubble expansion rate, $D_M(z)$ the comoving angular diameter distance, and $r_d$ the sound horizon at the drag epoch; quantities labeled ``fid'' are evaluated in the fiducial cosmology used to construct the galaxy catalogs and the two‑point correlation function (2PCF) template. Equivalently, $q_\parallel$ can be expressed via the radial Hubble distance $D_H(z) \equiv c/H(z)$ as $q_\parallel = (D_H/r_d)/(D_H^{\rm fid}/r_d^{\rm fid})$. Importantly, the dilation parameters modulate only the oscillatory BAO feature and are decoupled from the smooth broadband component of the correlation function.

For interpretational convenience, these are often reparameterized into an isotropic volume dilation $q_{\rm iso}$ and an Alcock--Paczynski anisotropy parameter $q_{\rm ap}$:
\begin{equation}
    q_{\rm iso}(z) \equiv \bigl[q_\parallel(z)\,q_\perp^2(z)\bigr]^{1/3}, \qquad
    q_{\rm ap}(z) \equiv \frac{q_\parallel(z)}{q_\perp(z)}.
\end{equation}

The position of the BAO peak in the reconstructed galaxy 2PCF is modeled by fitting its monopole and quadrupole moments, while marginalizing over the broadband shape using the spline-based technique described in \cite{KP4s2-Chen}. Our implementation follows the official DESI configuration‑space pipeline \citep{DESI.DR2.BAO.cosmo,DESI.DR2.BAO.lya} and adheres to the reconstruction conventions established in \cite{DESI2024.III.KP4,KP4s4-Paillas,KP4s3-Chen}.

We adopt the same scale cuts as the official DR1/2 BAO analyses, namely $s \in [50, 150]\,h^{-1}{\rm Mpc}$ with bin width $\Delta s = 4\,h^{-1}{\rm Mpc}$. The parameter vector $\Theta_{\rm BAO}$ may consist of $\{q_{\rm iso}^{\rm BAO}\}$, $\{q_{\rm iso}^{\rm BAO}, q_{\rm ap}^{\rm BAO}\}$, or $\{q_\parallel^{\rm BAO}, q_\perp^{\rm BAO}\}$, depending on the tracer's signal‑to‑noise and the adopted parameterization \citep{DESI.DR2.BAO.cosmo,DESI.DR2.BAO.lya,DESI2024.III.KP4}.

We sample the posterior distribution of BAO parameters using the \texttt{desilike} framework\footnote{\url{https://github.com/cosmodesi/desilike}}, which evaluates a multivariate Gaussian likelihood for the compressed BAO observables $\Theta_{\rm BAO}$:
\begin{equation}
    \log\mathcal{L}_{\rm BAO}(\Theta_{\rm BAO}) = -\frac{1}{2}\,
    \qty[\xi_{\rm data} - \xi_{\rm model}(\Theta_{\rm BAO})]^{T}
    \mathbf{C}_{\xi}^{-1}
    \qty[\xi_{\rm data} - \xi_{\rm model}(\Theta_{\rm BAO})],
\end{equation}
where $\xi_{\rm data}$ represents the measured monopole and quadrupole of the reconstructed 2PCF, $\xi_{\rm model}$ is the fiducial template rescaled by the dilation parameters, and $\mathbf{C}_{\xi}$ is the corresponding covariance matrix used in the DESI DR2 BAO analysis. We have verified that applying this pipeline to the DR2 EZmocks reproduces the official DESI DR2 BAO results, ensuring consistency with the collaboration's reference measurements.

\subsection{Full shape modeling}
In contrast to the primary DESI DR1 Full-Shape measurement \citep{DESI2024.V.KP5,DESI2024.VII.KP7B}, which relies on full  modeling of the galaxy power spectrum, we adopt  the ShapeFit \citep{Brieden21} model-agnostic compression technique. This approach parameterizes the power spectrum by projecting the full $k$-band information onto a compact set of compressed observables. Specifically, we extract scale dilation parameters $(q_{\rm iso}^{\rm SF}, q_{\rm ap}^{\rm SF})$,\footnote{We use the superscript ``SF'' to distinguish parameters derived from the full-shape ShapeFit analysis of pre-reconstructed catalogs from those labeled ``BAO,'' which are obtained by measuring only the BAO feature in the two-point correlation function of post-reconstructed catalogs. While covariant, these two measurements capture distinct cosmological information.} the logarithmic growth rate $f$ (often expressed as $df \equiv f / f_{\rm fid}$), two shape parameters $m$ and $n$ (expressed as $dm \equiv m - m_{\rm fid}$ and $dn \equiv n - n_{\rm fid}$), and an amplitude parameter $\sigma_{s8}$\footnote{This is the amplitude parameter $\sigma_{s8}^2 = \int_0^\infty P_{\rm lin}(k)$$W(8\,k\,s/(\Mpch))\dd^3k$, where $W(k)$ is a top-hat filter, $P_{\rm lin}$ is the Shapefit-modified linear power spectrum and $s = r_s / r_s^{\rm fid}$ denotes the change in the filter scale due to the scalings $q_{\rm iso,ap}$.} which describes the amplitude of the power spectrum. In the power spectrum-only analyses presented before, such as the companion paper \citep{desi1.5}, the growth and amplitude parameters are fully degenerate so only one of them is required for the cosmological interpretation (e.g. $f$). The inclusion of the bispectrum partially breaks this degeneracy, thus allowing us to extract $f$ and $\sigma_{s8}$ parameters separately  from the LRG bins where  the bispectrum measurements are included. For ELG and QSO we do not include the bispectrum:  systematics are difficult to model correctly and mitigate and, for QSO, the bispectrum is not constraining enough to disentangle the growth-amplitude degeneracy. The samples used in this analysis and the manner in which they are included is detailed in \cref{tab:desisamples}. For the  redshift bins where the bispectrum is not included, we keep only the $f$ parameter which is effectively reinterpreted as $f\sigma_{s8}$ and the SF likelihood is constructed as described in \cite{desi1.5} i.e., we use the power spectrum mono- and quadrupole in the range of $k\in \qty[0.02, 0.2]\hMpc$ in bins of $\Delta k = 0.005 \hMpc$. When including the bispectrum we follow \cite{novell-masot25}, thus the data vector becomes $PB$, the joint data vector including the mono- quad- and hexadecapole of the power spectrum and the bispectrum monopole in scales of $k \in [0.02, 0.15]\,h\mathrm{Mpc}^{-1}$ and $k \in [0.02, 0.12]\,h\mathrm{Mpc}^{-1}$ respectively, all with a binning of $\Delta k = 0.01\,h\mathrm{Mpc}^{-1}$. The likelihood is defined as 

\begin{equation}
    \log\mathcal{L}_{\rm SF}(\Theta_{\rm SF}) = -\frac{1}{2} \left[ PB_{\rm data} - PB_{\rm model}(\Theta_{\rm SF}) \right]^\mathrm{T} \mathbf{C}_{PB}^{-1} \left[ PB_{\rm data} - PB_{\rm model}(\Theta_{\rm SF}) \right],
\end{equation}
where $\Theta_{\rm SF} = \{q_{\rm iso}, q_{\rm ap}, dm, f, f\sigma_{s8}\}$ denotes the compressed parameter vector. The $dn$ parameter is kept fixed at 0 due to its large degeneracy with $dm$, which is interpreted as $dm + dn$ in the interpretation step \cite{novell-masot25,novell-masot26}. As in standard BAO analyses, additional nuisance parameters are marginalized over during inference. In particular, we assume coevolution for the $b_s$ bias parameter. We refer the reader to \cite{novell-masot25} for details regarding the treatment of nuisance parameters. The covariance matrix $\mathbf{C}_{PB}$ is the mock-estimated power spectrum-bispectrum joint covariance used in \cite{novell-masot25}, which includes the non-Gaussian contributions i.e. the $P(k)$-$B(k)$ cross covariance. Ignoring these off-diagonal blocks has been shown to underestimate the true error \cite{novell-masot24}.

The modeling of these statistics differs from the official DESI pipeline in that the power spectrum is computed using the Renormalized Perturbation Theory (RPT) instead of EFT and the hexadecapole of the power spectrum is added to the data vector. It has been shown in \cite{novell-masot25} that these two approaches yield consistent results in terms of compressed parameters and we refer the reader there for a thorough discussion on the matter. The bispectrum is modeled with the Geo-FPT approach \cite{geofpt}, which adopts the standard \lcdm{} perturbation kernels and is calibrated against \lcdm{} simulations, extending the tree-level bispectrum with a triangle shape-dependent factor that boosts the model accuracy up to $k\approx 0.12\hMpc$. We follow the pipeline implemented in \cite{novell-masot25}, including the priors in the SF and bias parameters, and we use the code \texttt{Brass}\footnote{\url{https://github.com/hectorgil/Brass.git}} to sample the SF posterior when including the bispectrum.

\subsection{Full-Shape Lyman-\texorpdfstring{$\alpha$}{alpha}}
At redshifts $z \gtrsim 2$, the Lyman-$\alpha$ (Ly$\alpha$) forest, i.e. the pattern of absorption imprinted on quasar spectra by neutral hydrogen in the intergalactic medium, emerges as the most statistically powerful tracer of large-scale structure accessible to DESI. While traditional analyses have focused on extracting cosmological information solely from the BAO peak position in the Ly$\alpha$ correlation function \cite{DESI.DR2.BAO.lya}, the full clustering signal encodes substantially more information. In particular, the broadband shape of the correlation function carries sensitivity to the Alcock-Paczynski (AP) effect, offering an independent geometric probe complementary to the BAO scale \cite{Cuceu2021,Cuceu2023}.

Building on this, \cite{Lya-fs} performed a full-shape analysis of the Ly$\alpha$ forest three-dimensional correlation functions, jointly modeling the BAO feature and the broadband clustering. By leveraging AP information extracted from the broadband shape in addition to that encoded in the BAO peak position, they achieved a 2.4-fold improvement in the precision of the distance ratio $D_H/D_M$ at the effective redshift $z_{\rm Ly\alpha}=2.33$, relative to a BAO-only analysis of the same dataset. This approach, which we denote as Ly$\alpha$ FS, thus combines geometric constraints from both the oscillatory and smooth components of the correlation function.

The analysis of Ref.~\cite{Lya-fs} provides combined constraints using DR1 full-shape data together with DR2 BAO measurements. For use in our joint cosmological inference, we adopt a Gaussian approximation to their likelihood, characterized by the mean vector and covariance matrix
\begin{equation}
    \boldsymbol{\mu}_{\rm Ly\alpha\text{-}FS} = 
    \begin{pmatrix}
        D_H(z_{\rm eff}) / r_{d} \\[2pt]
        D_M(z_{\rm eff}) / r_{d}
    \end{pmatrix}
    = 
    \begin{pmatrix}
        8.646 \\[2pt]
        38.90
    \end{pmatrix},
\qquad
    \mathbf{C}_{\rm Ly\alpha\text{-}FS} = 
    \begin{blockarray}{ccc}
     & D_H/r_{d} & D_M/r_{d} \\
    \begin{block}{c(cc)}
    D_H/r_{d} & 0.077 & -0.016 \\
    D_M/r_{d} & -0.016 & 0.38 \\
    \end{block}
    \end{blockarray},
\end{equation}
where $z_{\rm eff}$ denotes the effective redshift of the Ly$\alpha$ sample. This compressed likelihood encapsulates the key geometric constraints from the Ly$\alpha$ forest while remaining computationally efficient for combination with other DESI probes.

\subsection{Combination of DESI DR2 BAO and DR1 Full-Shape including the Bispectrum}
\label{sec:methods_combination}

Following the process outlined in \cite{desi1.5}, we use measurements performed on the mocks to obtain matrices ${\bf M}^{\rm BAO} = [\bf{q}^{\rm BAO}_{\rm iso}, \bf{q}^{\rm BAO}_{\rm ap}]$, which is of size $(N_{\rm mocks}\times N_{\rm params}) = (100 \times 2)$ for BAO; and ${\bf M}^{\rm SF} = [{\bf q}^{\rm SF}_{\rm iso}, {\bf q}^{\rm SF}_{\rm ap}, {\bf dm}, {\bf f}, {\bf f\sigma_{s8}}]$ (size $100 \times 5$). These are then concatenated into a matrix ${\bf M}^{\rm BAO + SF} = [{\bf q}^{\rm SF}_{\rm iso}, {\bf q}^{\rm SF}_{\rm ap}, {\bf dm}, {\bf f}, {\bf f\sigma_{s8}}, \bf{q}^{\rm BAO}_{\rm iso}, \bf{q}^{\rm BAO}_{\rm ap}]$ of size $100\times 7$. We can then use these measurements matrices to compute $7\times7$ covariance matrices ${\bf C}^{\rm X} = (\Delta{\bf M}^{\rm X})^{T}(\Delta{\bf M}^{\rm X})$, where $\Delta{\bf M}^{\rm X}\equiv {\bf M}^{\rm X} - \bar{{\bf M}}^{\rm X}$ is the matrix of residuals with $\bar{{\bf M}}^{\rm X}$ the ensemble-mean of measurements. In the case of $\mathbf{C}^{\rm BAO + SF}$, we divide it in three parts as
\begin{equation}
     \mathbf{C}^{\rm BAO + SF} = \begin{pmatrix}
   \mathbf{C}^{\rm SF} & (\mathbf{C}^{\rm BAO \times SF})^{T}\\\
\mathbf{C}^{\rm BAO \times SF} & \mathbf{C}^{\rm BAO}
 \end{pmatrix},
 \label{eq:block_matrix}
\end{equation}

What is actually extracted directly from the mocks are the associated correlation matrices $(\mathbf{r}^X)_{ij}  = (\mathbf{C}^X)_{ij} / \sqrt{(\mathbf{C}^X)_{ii}(\mathbf{C}^X)_{jj}}$, thus :
\begin{equation}
     \mathbf{r}^{\rm BAO + SF} = \begin{pmatrix}
   \mathbf{r}^{\rm SF} & (\mathbf{r}^{\rm BAO \times SF})^{T}\\\
\mathbf{r}^{\rm BAO \times SF} & \mathbf{r}^{\rm BAO}
 \end{pmatrix}.
 \label{eq:block_matrix_correlation}
\end{equation}

The diagonal blocks are estimated by measuring the matrices ${\mathbf{r}}^{\rm BAO}_{\rm data}$, ${\mathbf{r}}^{\rm SF}_{\rm data}$ from the posterior samples of the official BAO and ShapeFit measurements from the DESI data and from \citep{novell-masot25} for the samples including the bispectrum.

Let
    \begin{equation}
             \mathbf{r}^{\rm BAO + SF}_{\rm mock} = \begin{pmatrix}
   \mathbf{r}^{\rm SF}_{\rm mock} & (\mathbf{r}^{\rm BAO \times SF}_{\rm mock})^{T}\\\
\mathbf{r}^{\rm BAO \times SF}_{\rm mock} & \mathbf{r}^{\rm BAO}_{\rm mock}
 \end{pmatrix},
    \end{equation}
    be the correlation matrix computed from the mocks, i.e. associated to the covariance matrix computed from the mock measurements in compressed variable space
    \begin{equation}
       \mathbf{C}^{\rm BAO + SF}_{\rm mock} \equiv (\Delta\mathbf{M}^{\rm BAO + SF})^T(\Delta\mathbf{M}^{\rm BAO + SF}).
    \end{equation}
Following \cite{desi1.5} we use the diagonal blocks obtained from the data posteriors in the individual analyses and estimate the off-diagonal blocks from the mocks.
    \begin{equation}
             \mathbf{r}^{\rm BAO + SF}_{\rm comb} = \begin{pmatrix}
   \mathbf{r}^{\rm SF}_{\rm data} & (\mathbf{r}^{\rm BAO \times SF}_{\rm mock})^{T}\\\
\mathbf{r}^{\rm BAO \times SF}_{\rm mock} & \mathbf{r}^{\rm BAO}_{\rm data}
 \end{pmatrix}.
 \label{eq:matrix_combination}
    \end{equation}
    
This correlation matrix is then converted back into a covariance using the data variance as
\begin{equation}
\label{eq:combined_covariance}
    (\mathbf{\hat{C}}^{\rm BAO + SF}_{\rm comb})_{ij} = \sqrt{(\mathbf{C}^{\rm BAO + SF}_{\rm data})_{ii}(\mathbf{C}^{\rm BAO + SF}_{\rm data})_{jj}}(\mathbf{r}^{\rm BAO + SF}_{\rm comb})_{ij}.
\end{equation}

Though $\mathbf{r}^{\rm BAO + SF}_{\rm comb}$ is our baseline and recommended approach, we also quantify the effect of ignoring these cross-covariance blocks by repeating some of the analyses using a correlation 
\begin{equation}
             \mathbf{r}^{\rm BAO + SF}_{\rm data} = \begin{pmatrix}
   \mathbf{r}^{\rm SF}_{\rm data} & \mathbf{0}\\\
\mathbf{0} & \mathbf{r}^{\rm BAO}_{\rm data}
 \end{pmatrix}.
 \label{eq:matrix_combination_zeros}
    \end{equation}

The combined analysis follows the standard approach \cite{desi1.5, DESI2024.VII.KP7B}.
The likelihood is built as a multivariate Gaussian for the covariance matrix of \cref{eq:matrix_combination} for LRG and for the matrices provided by \cite{desi1.5} for ELG and QSO.
\begin{align}
    \log\mathcal{L}_{\rm LEQ} =& \sum_{i \in~ \rm LRG,ELG,QSO}\log\mathcal{L}_{i}(z_{\rm eff}, \Theta_{\rm cosmo}) \\=& -\frac{1}{2}\sum_{i \in \rm LRG,ELG,QSO}\qty[\hat{\Theta}_{i,\rm comp}^{*,\,\rm data} - \Theta_{\rm comp}^{\rm model}]^T\qty(\mathbf{\hat{C}}^{\rm BAO + SF}_{\rm comb})_i^{-1}\qty[\hat{\Theta}_{\rm comp}^{*,\,\rm data} - \Theta_{i,\rm comp}^{\rm model}],
\end{align}
where the compressed vector $\Theta_{\rm comp} = \qty[\Theta_{\rm SF},\Theta_{\rm BAO}]$ is the concatenation of BAO and ShapeFit compressed parameter vectors and the $\Theta^{\rm model}_{\rm comp}$ is evaluated at the point $\Theta_{\rm cosmo}$ in the cosmological parameter space. The superscript $(*,\rm data)$ denotes the maximum a posteriori point in compressed parameter space extracted from the data posterior for each sample.

We include BGS data only in BAO form, as we have not generated DR2 BGS EZmocks. Similarly, we take the Lyman-$\alpha$ forest into account using the DR2 BAO + DR1 Full-Shape likelihood $\mathcal{L}_{\rm Ly\alpha-FS}$. We combine these into the DESI likelihood
\begin{equation}
    \log\mathcal{L}_{\rm DESI} \equiv \log\mathcal{L}_{\rm LEQ} + \log\mathcal{L}_{\rm BGS} + \log\mathcal{L}_{\rm Ly\alpha-FS}.
\end{equation}

For cosmological inference, we employ the \texttt{Cobaya} \cite{Cobaya} framework with \texttt{CAMB} \cite{CAMB} and use the external likelihoods implemented in the package when appropriate. 

\section{Results and Discussion}
\label{sec:results}

\subsection{Covariance matrices in compressed space}
\label{sec:results_combination}

\begin{figure}
    \centering
    \includegraphics[width=0.99\linewidth]{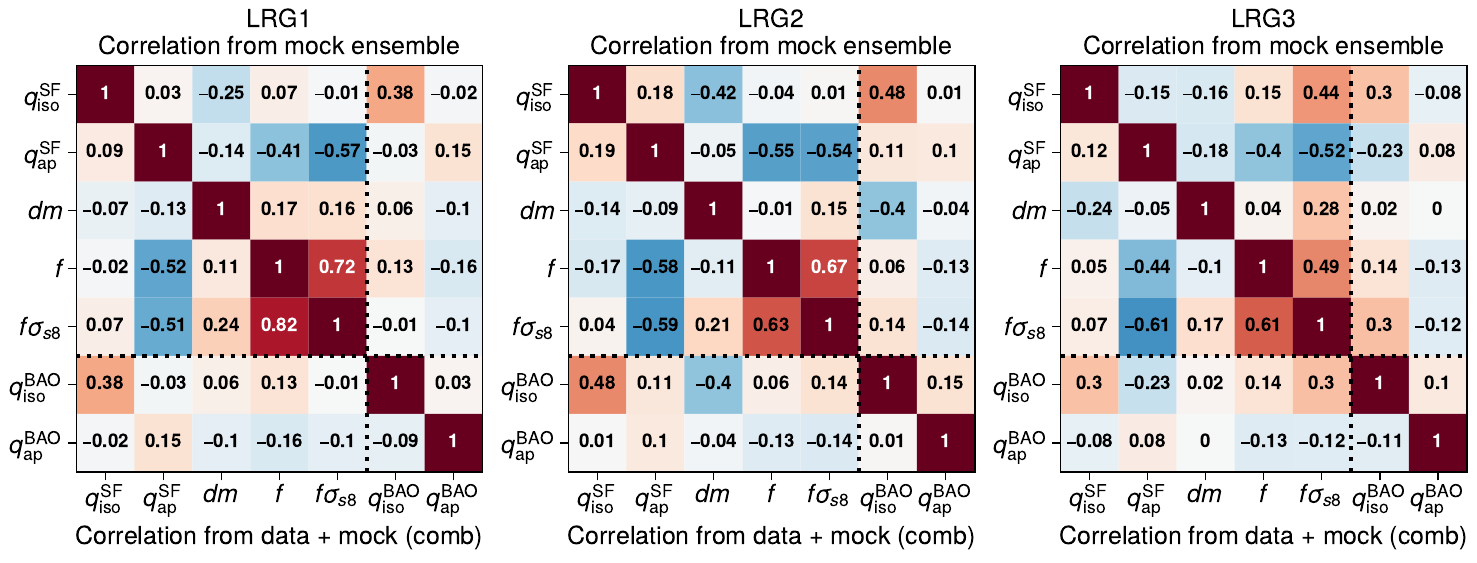}
    \caption{DESI1.5 + Bispectrum correlation matrices for the LRG samples used in this work. We show the matrices divided in the structure shown in \cref{eq:block_matrix_correlation} with dotted lines separating the SF, BAO and cross blocks. The upper triangles show the correlations obtained from the mocks, while the lower triangles show the combined correlations, where the auto blocks (BAO $\times$ BAO and SF $\times$ SF)  matrices come from the data posteriors and the off-diagonal cross (BAO $\times$ SF) blocks come from the mocks. Colors show the degree of correlation ranging from dark blue for complete anti-correlation (-1) to complete correlation in dark red (1). The correlation values themselves are also shown.}
    \label{fig:covariances}
\end{figure}

Our estimated correlation matrices are shown in \cref{fig:covariances}, where we highlight the structure introduced in \cref{eq:block_matrix_correlation}. Values for elements below the diagonal correspond to $\mathbf{r}_{\rm comb}$, (our baseline) and values for elements  above the diagonal correspond to $\mathbf{r}_{\rm mock}$.  The off-diagonal cross BAO$\times$SF  blocks are identical as they come from the mocks in both cases, whereas the off-diagonal values within the auto BAO$\times$BAO and FS$\times$SF blocks differ. The structure of these matrices is consistent, suggesting that the mock covariance is a fair representation of the statistics of the data.

The variances used in our covariance matrix (not shown in the figure) are taken directly from the data. For the BAO block, we use the variances from the official DESI BAO posteriors. For the FS block, we use the variances from the posteriors of \cite{novell-masot25}, which already incorporate the systematic error budget estimated in that work. To remain consistent with the procedure and systematic error budget of \cite{novell-masot25}, we adopt their choice of parameterization for the compressed SF variables, namely $f$ and $f\sigma_{s8}$. We have tested alternative parameterizations (e.g., $\sigma_{s8}$ together with $f\sigma_{s8}$) and found that they yield equivalent cosmological constraints.

\begin{figure}
    \centering
    \includegraphics[width=0.75\linewidth]{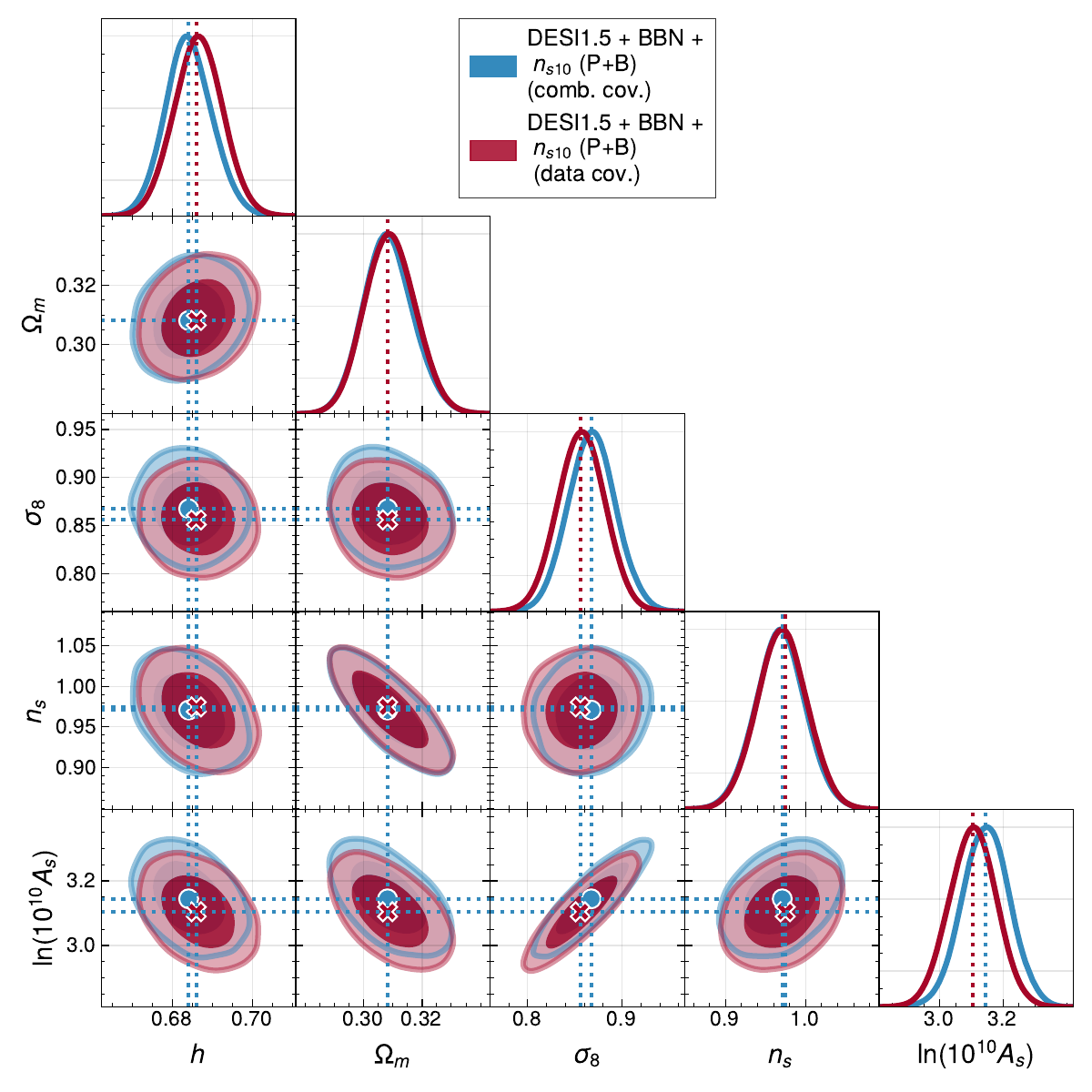}
    \caption{Triangle plot showing our baseline \lcdm{} results with the combined covariance (i.e including the cross covariance) and the corresponding constraints when ignoring the cross-covariance blocks (i.e. the data-only covariance).}
    \label{fig:cross-cov}
\end{figure}

In \cite{desi1.5} we found that in compressed variable space and for power spectrum information only,  ignoring the covariance between DR1 FS and DR2 BAO  does not affect  cosmological inference in any significant way. For completeness, we repeat these tests while including the bispectra. \Cref{fig:cross-cov} compares the \lcdm{} constraints with and without the cross-covariance blocks $\mathbf{r}^{\rm BAO\times SF}$. The absence of these blocks in the covariance results in a $\sim 0.44\sigma$ shift in $\sigma_8$ ($0.55\sigma$ shift in $\log(10^{10}A_s)$) and a $0.4\sigma$ shift in $h$, but does not affect the overall constraining power.

This larger impact of the cross-covariance blocks relative to the power spectrum-only analyses can be attributed to the extra constraining power in the bispectrum and the new covariance structure derived from the amplitude-growth degeneracy breaking. First, the SF projection with the bispectrum splits the growth and amplitude directions, allowing each to individually correlate with the BAO information from DR2. Second, the BAO information from SF encoded in the dilation parameter $q_{\rm iso}$ is measured to be more correlated with the DR2 BAO than in the power spectrum-only case (see \cref{fig:covariances}; cf. Fig.~1 of \cite{desi1.5} for the power-spectrum-only correlations). This new correlation structure, especially in the amplitude-growth sector, should be expected to affect cosmological measurements, in particular the amplitude parameters, more than the power spectrum counterpart, where these directions were completely degenerate.

Additionally, the DR1 data is a strict subset of the DR2, meaning that the cross correlation cannot be zero by construction, which is our main reason to keep the cross-covariance block in our analysis.

 Having 100 mocks ($N_{\rm Mocks}$) to estimate the covariance among $n=7$ parameters places us in the regime $N_{\rm Mocks}\gg n$, where the effects of the finite number of mocks on the cosmological inference are perturbative. In particular there are several well-known effects  and corrections at play, which we briefly enumerate.
At first order, the finite number of mocks biases the inverse covariance; this is corrected approximately by the Hartlap factor \cite{Hartlap2007} and exactly by adopting a Sellentin--Heavens likelihood~\cite{SellentinHeavens2016}. At second order, residual precision-matrix noise from the finite $N_{\rm Mocks}$ \cite{DodelsonSchneider2013} causes the recovered best-fit parameters to scatter from realization to realization, with a corresponding effect on posterior widths that depends on the number of parameters being estimated and is quantified by \cite{Percival2014}. Both effects scale as $n/N_{\rm Mocks}$ in the parameter variance, i.e.\ as $(n/N_{\rm Mocks})^{1/2}$ in units of the parameter uncertainty, and are therefore not necessarily negligible.
Because our covariance matrix is hybrid\new{,} the assumptions underlying the theoretical predictions may not strictly hold (e.g. the distribution is not Wishart) and the analytical estimates may not be fully correct. 
We thus proceed as follows.

We estimate the finite Mocks sampling errors in the cross-covariance blocks by bootstrapping the covariance matrices.  These errors are of order 0.1 (\cref{app:cross-uncert}, \cref{fig:correlation_errors} and \cref{tab:cross_corr}), consistent with the expected sampling error of a correlation coefficient computed from $N_{\rm Mocks}=100$ realisations. We propagate their effect on the cosmological parameters by importance sampling (see \cref{app:robustness_cosmo} for details).  We add the resulting scatter of the bootstrap posterior means, $\mathrm{std}_b(\mu_b)$ (the analogue of the effect identified by \cite{DodelsonSchneider2013}), in quadrature to the errors of every quoted parameter. We quote the total uncertainty as
$\sigma_{\mathrm{quad}} \equiv \sqrt{\sigma_{\mathrm{ref}}^2 + \mathrm{std}_b(\mu_b)^2}$.
We refer to the extra contribution $\mathrm{std}_b(\mu_b)$ as the finite-mock error. This is the more conservative approach as it inflates the errors by default while approximating the Bayesian marginal error at fixed data, $\langle\sigma_b^2\rangle+\mathrm{Var}(\mu_b)$, when our statistical error $\sigma_{\mathrm{ref}}$ approaches the mean error $\langle\sigma_b^2\rangle$. In practice, including it increases the quoted uncertainties by at most ${\approx}10\%$, and typically by a few per cent.

 The estimate is performed independently for every cosmological model and dataset combination considered in this work. For $\sum m_\nu$ in the neutrino-mass extensions (\nulcdm{}, \nulcdm{}+CMB and \nuwowacdm{}), whose posteriors are truncated at zero, we propagate the finite-mock uncertainty to the 95\% CL by marginalizing over the bootstrap realizations, i.e. we compute the shift
\begin{equation}
\begin{split}
  \Delta_{95} &= Q_{95}\!\qty[\bar{P}\qty(\textstyle\sum m_\nu)]
               - Q_{95}\!\qty[P\qty(\textstyle\sum m_\nu \,\middle|\,
                 \mathbf{\hat{C}}^{\rm BAO+SF}_{\rm comb})]\,, \\
  \bar{P}\qty(\textstyle\sum m_\nu)
              &\equiv \frac{1}{N_b}\sum_{b=1}^{N_b}
                 P\qty(\textstyle\sum m_\nu \,\middle|\,
                 \mathbf{\hat{C}}^{\rm BAO+SF}_{{\rm comb},b})\,,
\end{split}
\label{eq:delta95}
\end{equation}
where $P(\sum m_\nu \,|\, \mathbf{\hat{C}})$ is the posterior of $\sum m_\nu$ obtained under the covariance $\mathbf{\hat{C}}$, marginalized over all other parameters, $Q_{95}[\cdot]$ denotes the 95th percentile of the bracketed distribution, and $\hat{C}^{\rm BAO+SF}_{{\rm comb},b}$ is the $b$-th bootstrap realization (\cref{app:cross-uncert}), with $b = 1, \dots, N_b$ and $N_b = 500$--$2000$ depending on the dataset\footnote{We use 500 mocks for chains with CMB, which are longer, and 2000 for all other cases.}. The averaged density $\bar{P}$ is thus the posterior further marginalized over the covariance uncertainty: the 95\% CL quoted in \cref{sec:nucdm_constraints} corresponds to the largest among $Q_{95}[\bar{P}]$ and $Q_{95}\!\qty[P\qty(\textstyle\sum m_\nu \,\middle|\,
\mathbf{\hat{C}}^{\rm BAO+SF}_{\rm comb})]$, that is, when $\Delta_{95}$ is negative, we keep the original limit to be more conservative.


\subsection{Cosmological inference}

\begin{figure}[t]
    \centering
    \includegraphics[width=0.49\linewidth]{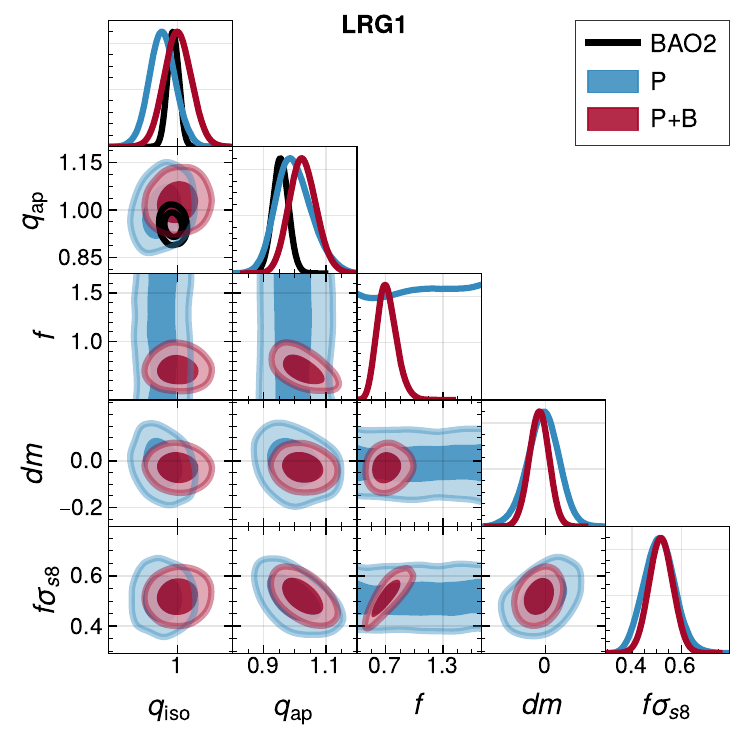}
    \includegraphics[width=0.49\linewidth]{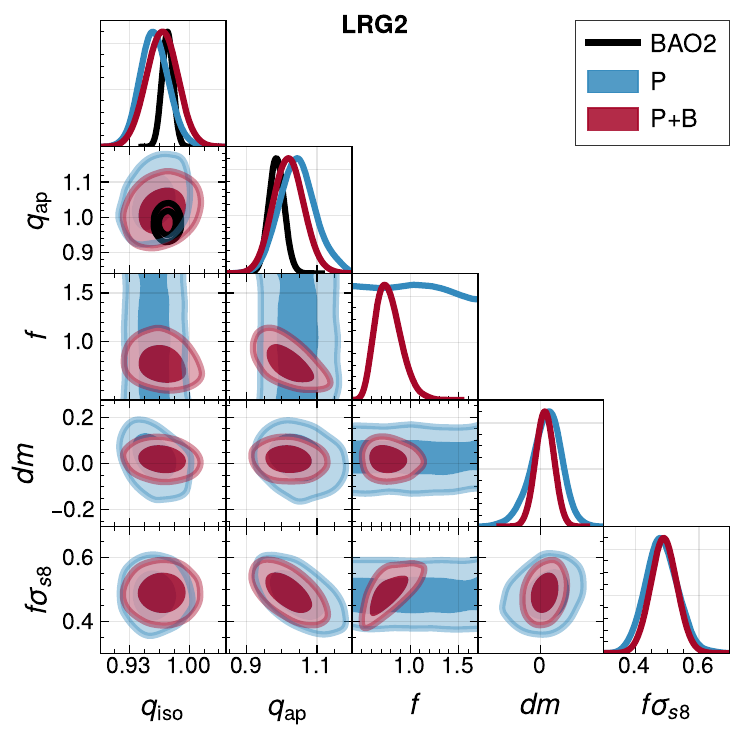}
    \includegraphics[width=0.49\linewidth]{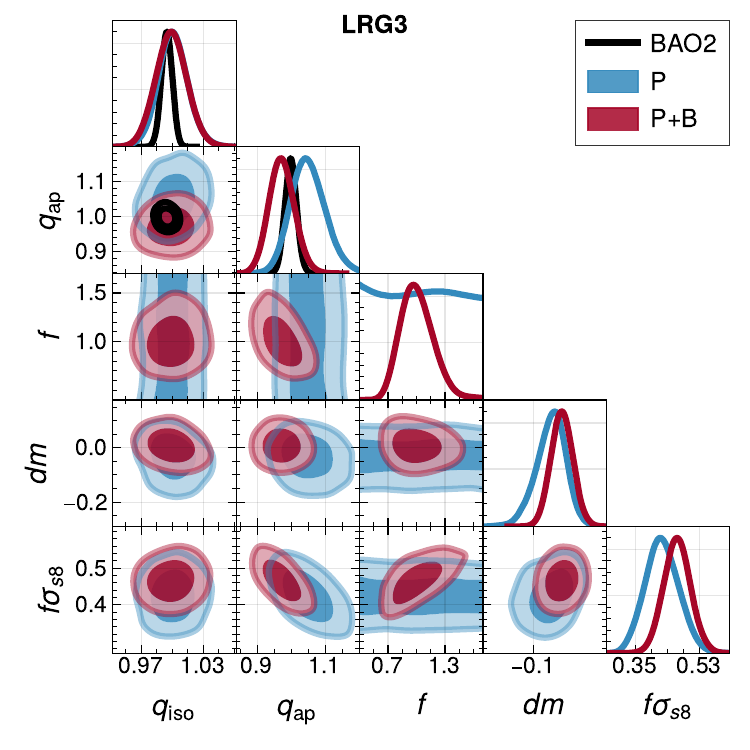}
    \caption{Comparison of 2-point (P) and 3-point (P+B)-based Shapefit constraints for the different DESI DR1 LRG bins. The 2-point measurements are those of Ref.~\cite{DESI2024.V.KP5} and the 3-point are from Ref.~\cite{novell-masot25}. Since the power spectrum chains were run with a single degenerate parameter, we only show constraints for $f\sigma_{s8}$ and a flat posterior for $f$. On the other hand, for the bispectrum, both $\sigma_{s8}$ and $f$ are sampled independently thus we report both. We have added the contours for BAO2 as well for reference.}
    \label{fig:sf-lrgs}
\end{figure}

\begin{figure}
    \centering
    \includegraphics[width=0.99\linewidth]{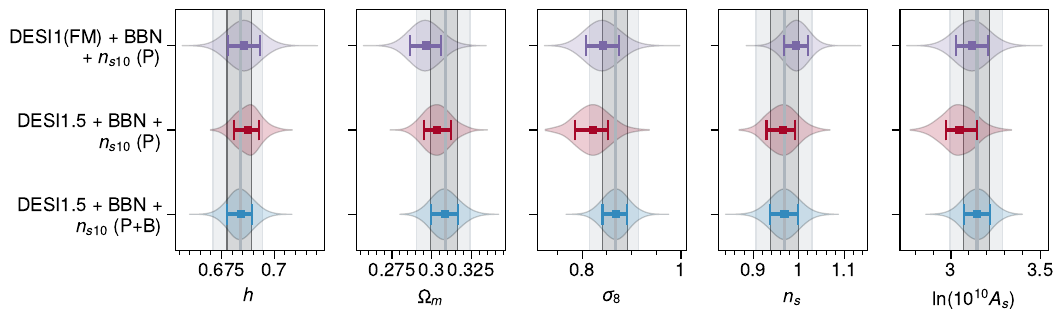}
    \caption{Comparison of our results (DESI1.5 + Bispectrum, labeled P+B) to previous DESI1 and DESI1.5 analyses (using only the power spectrum, labeled ``P'') in the context of \lcdm. Constraints from DESI1 using FM from \cite{DESI2024.VII.KP7B} shown for reference.}
    \label{fig:lcdm}
\end{figure}
\begin{table}
    \centering
    \resizebox{\columnwidth}{!}{\begin{tabular}{p{3cm}ccccc}
        \toprule
        Dataset & $h$ & $\Omega_m$ & $\sigma_8$ & $n_s$ & $S_8$\\
\midrule 
\rowcolor[HTML]{EFEFEF}
\textbf{DESI1.5 + BBN + $n_{s10}$ (P+B)} & \boldmath$0.6841 \pm 0.0060$  & \boldmath$0.3085 \pm 0.0087$  & \boldmath$0.868 \pm 0.026$  & \boldmath$0.969 \pm 0.031$  & \boldmath$0.880 \pm 0.028$ \\ 
DESI1.5 + BBN + $n_{s10}$ (P) & $0.6876 \pm 0.0059$  & $0.3035 \pm 0.0085$  & $0.822 \pm 0.034$  & $0.965 \pm 0.032$  & $0.827 \pm 0.036$ \\ 
        \bottomrule
                        \end{tabular}}
                        \caption{Comparison of the present DESI1.5 (P+B) \lcdm{} constraints with the previous power-spectrum-only DESI1.5 results of \cite{desi1.5}. The (P+B) analysis yields higher central values and $\sim$24\% and $\sim$22\% tighter constraints on $\sigma_8$ and $S_8$, respectively. Since the two full-shape analyses are not matched in all respects, the differences between the two rows should not be interpreted solely as the incremental effect of adding the bispectrum; a controlled DR1 (P) versus (P+B) comparison is presented in \cite{novell-masot25,novell-masot26}. We highlight in boldface the constraints new to this work\protect\footnotemark.}
                        \label{tab:lcdm_tab}
                    \end{table}

\footnotetext{$S_8$ is not sampled in the IS runs: its total error ($\pm0.028$) is obtained by propagating the $\sigma_8$ inflation factor ($\sigma_{\rm quad}/\sigma_{\rm ref} = 1.067$) to the quoted statistical error; all other errors are the totals $\sigma_{\rm quad}$.}
We first show the SF constraints for both power spectrum \cite{DESI2024.V.KP5} and bispectrum \cite{novell-masot25} in \cref{fig:sf-lrgs}. The plot shows that in the case of the bispectrum, we are able to independently constrain $f$ and $\sigma_{s8}$, while the power spectrum chains are run only using a single degenerate parameter $f\sigma_{s8}$. No power spectrum chains were run sampling growth and amplitude independently since the 2-point clustering can't disentangle this degeneracy so we show a band for $f$ in this case. In addition we show the BAO constraints from \cite{DESI.DR2.BAO.cosmo} to show they are consistent with both P and P+B geometry constraints. We expect some differences between the P and P+B $q_{\rm ap}$ parameter as the P+B results shown also include the hexadecapole of the power spectrum. Nonetheless, given that the BAO2 geometry constraints are much tighter, these will dominate when combining with the SF constraints to obtain cosmological parameter estimates.

While  the addition of the bispectrum  breaks the $f-\sigma_{s8}$ degeneracy,
 in \lcdm{} $f$ is determined by $\Omega_m$. Hence in this case the bispectrum is only  expected to improve constraints on $\sigma_8$.

\subsubsection{\lcdm{} constraints}\label{sec:lcdm_constraints}
             
Our baseline results are obtained including the minimum external data possible, i.e. BBN ($\omega_b$) and wide $n_{s}$ priors as described in \cref{sec:external}. \Cref{fig:lcdm} shows the \lcdm{} constraints (also summarized in \cref{tab:lcdm_tab}) obtained from the joint analysis (bottom line) as well as our previous DESI1.5 measurements in the middle row and the official DESI analysis in the top row.

\begin{equation}
\label{eq:lcdm_constraints}
    \begin{rcases}
    \Omega_m &= \text{$0.3085 \pm 0.0087$} \\
\sigma_8 &= \text{ $0.868 \pm 0.026$} \\
h &= \text{ $0.6841 \pm 0.0060$} \\
S_8 &= \text{$0.880 \pm 0.028$} \\
\end{rcases}
    \quad
    \text{DESI1.5 + BBN + $n_{s10}$ (P+B)} 
    \end{equation}

\Cref{eq:lcdm_constraints} shows the final constraints obtained when including the LRG bispectrum. Compared with the previous power-spectrum-only DESI1.5 analysis of \cite{desi1.5}, the present (P+B) analysis yields a higher clustering amplitude, with $\sigma_8$ changing from $0.822\pm0.034$ to  $0.868\pm0.026$ and $S_8$ from $0.827\pm0.036$ to $0.880\pm0.028$. The marginalized uncertainties are reduced by 23.5\% and 22.2\%, respectively. We stress that this numerical comparison does not constitute a strictly controlled estimate of the incremental effect of the bispectrum, since the power-spectrum-only and (P+B) full-shape inputs are taken from different analyses and are not identical apart from the inclusion of the bispectrum (B). A matched comparison using the same DR1 analysis framework was performed in \cite{novell-masot26}, where the addition of the bispectrum was shown to break the growth-amplitude degeneracy and improve the constraints on amplitude parameters. The present results show how the (P+B) information propagates when combined with DR2 BAO, including the corresponding cross-covariance. Similarly high values of $\sigma_8$ have also been reported from DESI data within the frequentist analysis of DR1 FS \cite{Morawetz-frequentist}. 

The improvements in the amplitude constraints are consistent with the findings of \cite{novell-masot26} using only DR1 data, where the constraint on amplitude parameter $\log(10^{10}A_s)$ also improved by roughly $20\%$ due to the addition of the bispectra. These larger $S_8$ values are in a $\sim3\sigma$ tension with the DES Y6 results \cite{DESY6GG,DESY6} but are consistent ($\sim1\sigma$) with the CMB-only constraints \cite{ACTDR6,SPT-3G}.

In contrast, the background parameters $h$, $\Omega_m$ and the spectral index $n_s$ show only marginal shifts (less than $2\%$) and no significant change in their uncertainties as these are dominated by the BAO2 information already present in the DESI1.5 (P) analysis.

\subsubsection{\wowacdm{} Dynamical Dark Energy}

\label{sec:w0wacdm_constraints}

\begin{table}
    \centering
    \resizebox{\columnwidth}{!}{\begin{tabular}{p{3cm}cccccc}
        \toprule
        Dataset & $h$ & $\Omega_m$ & $\sigma_8$ & $n_s$ & $w_0$ & $w_a$\\
\midrule 
\rowcolor[HTML]{EFEFEF}
DESI1.5 +  BBN + 
$n_{s10}$ (P) & $0.645 \pm 0.022$  & $0.351 \pm 0.025$  & $0.782 \pm 0.038$  & $0.956 \pm 0.039$  & $-0.49 \pm 0.25$  & $-1.52 \pm 0.77$ \\ 
\textbf{DESI1.5 + BBN + 
$n_{s10}$ (P+B)} & \boldmath$0.651 \pm 0.023$  & \boldmath$0.342 \pm 0.026$  & \boldmath$0.838 \pm 0.034$  & \boldmath$0.972 \pm 0.041$  & \boldmath$-0.64 \pm 0.26$  & \boldmath$-1.04 \pm 0.82$ \\ 
\rowcolor[HTML]{EFEFEF}
DESI1.5 + CMB (P) & $0.634 \pm 0.017$  & $0.356 \pm 0.020$  & $0.778 \pm 0.015$  & $0.9667 \pm 0.0035$  & $-0.38 \pm 0.20$  & $-1.82_{-0.53}^{+0.66}$ \\ 
\textbf{DESI1.5 + CMB (P+B)} & \boldmath$0.640 \pm 0.018$  & \boldmath$0.351 \pm 0.020$  & \boldmath$0.789 \pm 0.015$  & \boldmath$0.9655 \pm 0.0036$  & \boldmath$-0.48 \pm 0.20$  & \boldmath$-1.57 \pm 0.58$ \\ 
\rowcolor[HTML]{EFEFEF}
\textbf{DESI1.5 +
BBN + $n_{s10}$ + DES-Dovekie (P+B)} & \boldmath$0.673 \pm 0.010$  & \boldmath$0.3168 \pm 0.0091$  & \boldmath$0.859 \pm 0.027$  & \boldmath$0.977 \pm 0.040$  & \boldmath$-0.882 \pm 0.065$  & \boldmath$-0.34 \pm 0.30$ \\ 
\textbf{DESI1.5 +
DES-Dovekie + 
 CMB (P+B)} & \boldmath$0.6723 \pm 0.0056$  & \boldmath$0.3160 \pm 0.0057$  & \boldmath$0.8147 \pm 0.0077$  & \boldmath$0.9669 \pm 0.0036$  & \boldmath$-0.817 \pm 0.057$  & \boldmath$-0.68 \pm 0.21$ \\ 
        \bottomrule
                        \end{tabular}}
\caption{\wowacdm{} constraints for DESI1.5 with and without LRG bispectra. The inclusion of the bispectrum shifts the DE parameters closer to \lcdm{} and slightly improves $\sigma_8$. }
\label{tab:cosmo_params_w0wa}
\end{table}

When assuming a model with dynamical dark energy, we use the CPL \cite{ChevallierPolarski, Linder03} parametrization. Our previous work provided the first robust Bayesian DESI-only constraints on this model using the FS data. In this work we further build on those by adding the bispectrum of the LRG, while still considering both the cross correlation between power spectra and bispectra as well as the cross correlation between BAO2 and FS data using the ShapeFit approach. 

\Cref{tab:cosmo_params_w0wa} summarizes these constraints alongside a comparison with the previous DESI1.5 analyses. Similarly to the \lcdm{}, the additional three-point information shifts the $\sigma_8$ posterior mean from $0.782\pm 0.038$ to $0.838\pm 0.034$, accompanied by an $11\%$ reduction in uncertainty relative to the power-spectrum-only constraints. The dark energy equation-of-state parameters consistently migrate toward the \lcdm{} point, with $w_0$ moving from $-0.49\pm 0.25$ to $-0.64\pm 0.26$ and $w_a$ from $-1.52\pm 0.77$ to $-1.04\pm 0.82$.

\Cref{eq:w0wacdm_constraints} shows our baseline constraints on DE where the constraint on each parameter are marginalized over all other parameters:

    \begin{equation}
    \label{eq:w0wacdm_constraints}
    \begin{rcases}
w_0 &= \text{$-0.64 \pm 0.26$} \\
w_a &= \text{$-1.04 \pm 0.82$} \\
\end{rcases}
    \quad
    \text{DESI1.5 + BBN + $n_{s10}$  (P+B)}.
    \end{equation}

\begin{figure}
    \centering
    \includegraphics[width=0.49\linewidth]{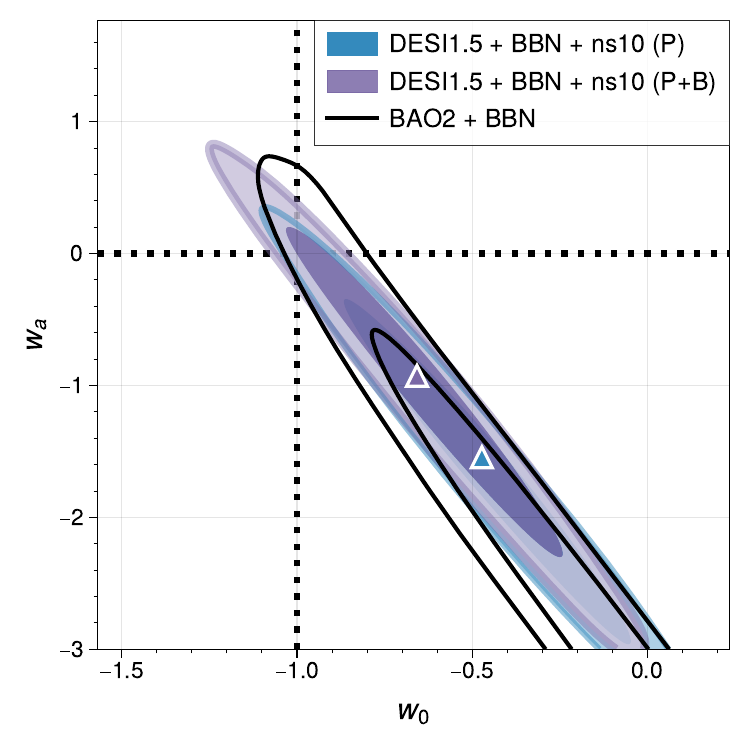}
    \includegraphics[width=0.49\linewidth]{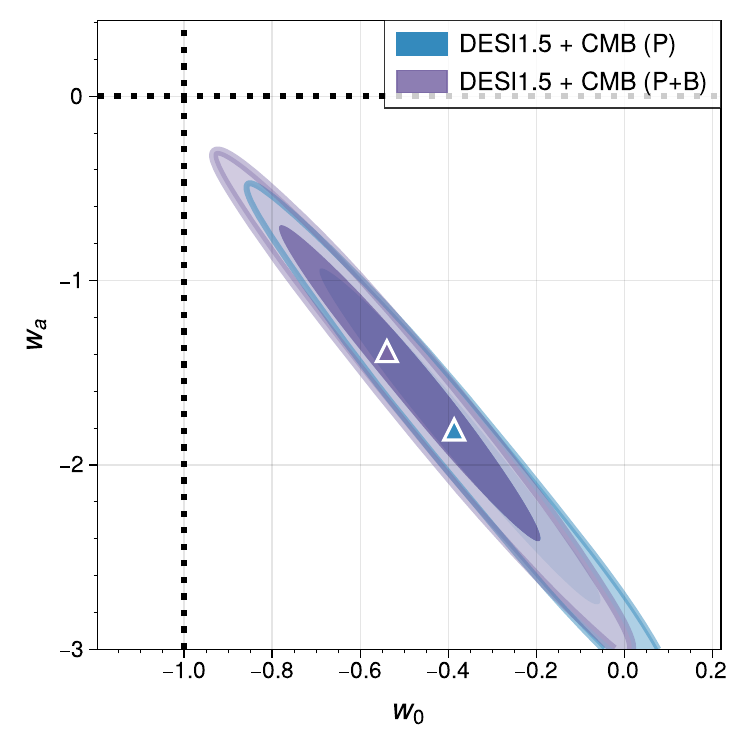}
    \includegraphics[width=0.49\linewidth]{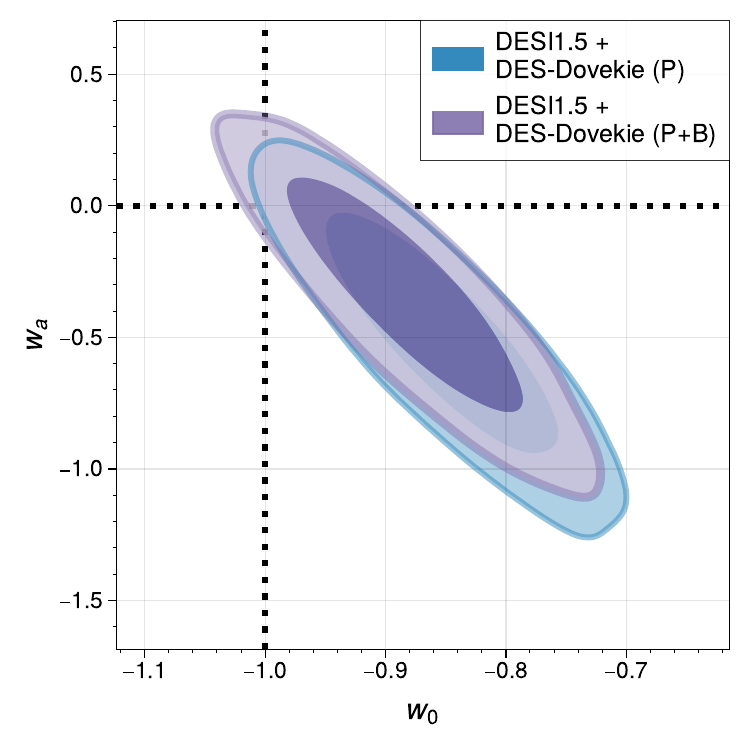}
    \includegraphics[width=0.49\linewidth]{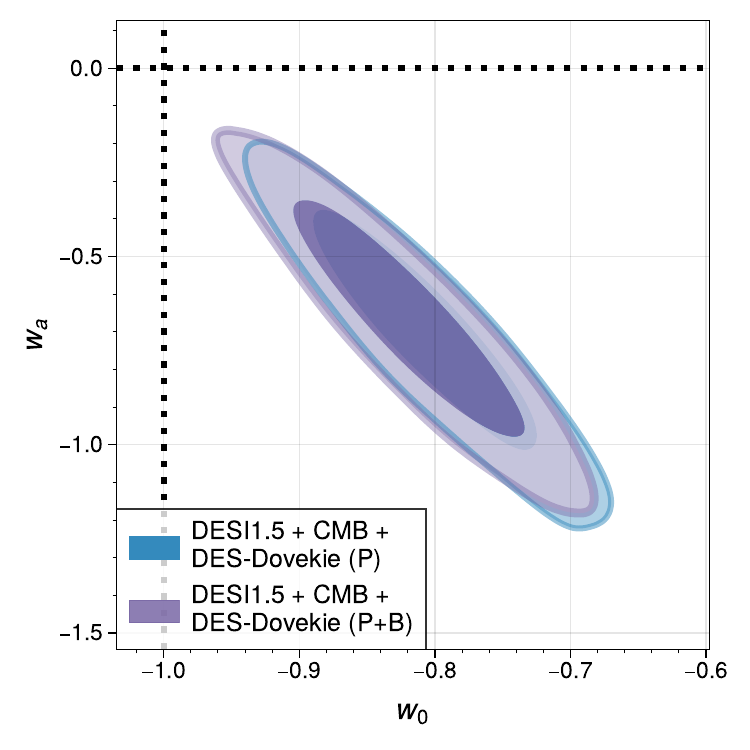}
    \caption{Dark energy equation-of-state constraints from DESI1.5 with and without the addition of the LRG bispectra. Top left: Parameter constraints with minimal external data. We have added the MAP points for the FS constraints to highlight the absence of projection effects, as well as the BAO2 constraint to highlight the information content of FS analyses. Top right: DE constraints adding CMB. Bottom left: Constraints only with SNe from DES. Bottom right: DE constraints combining all datasets.}
    \label{fig:w0wa}
\end{figure}

A closer examination of the $w_0$--$w_a$ plane (displayed in the top-left panel of \cref{fig:w0wa}) reveals that the inclusion of bispectrum information systematically shifts the joint confidence contours toward the $\Lambda$CDM point ($w_0=-1$, $w_a=0$) along the degeneracy direction. In the two-dimensional plane, the $\Lambda$CDM point lies inside the $1\sigma$ contour after adding the bispectrum, indicating consistency at that level. This might appear counterintuitive because the one-dimensional marginalised constraints reported in \cref{eq:w0wacdm_constraints} exclude the $\Lambda$CDM values by more than $1\sigma$ (e.g., $w_0 = -0.64\pm0.26$ is $1.4\sigma$ away from $-1$). The resolution of this apparent contradiction lies in the strong negative correlation between $w_0$ and $w_a$ inherent to the CPL parametrisation. The one-dimensional posteriors project the correlated two-dimensional contour onto each axis, thereby amplifying the deviation from $\Lambda$CDM in the marginalised distributions even when the joint posterior comfortably contains the $\Lambda$CDM point.

To underscore the robustness of this inference against prior-induced projection effects, we overlay the maximum a-posteriori (MAP) estimates for each analysis variant. This diagnostic helps distinguish genuine shifts in the likelihood peak from artifacts arising from marginalization over poorly constrained nuisance directions, which is a known concern in high-dimensional Bayesian analyses, where residual volume effects can displace contours along degeneracy trajectories even when alternative mitigation strategies are employed. We have also added the BAO2 contours to highlight the wealth of information present in the FS analysis.

These represent the most up-to-date and stringent DESI-only constraints (supplemented only by minimal external priors) on dynamical dark energy to date, showing no statistically significant preference for time-evolving dark energy. 

In the top right panel of \cref{fig:w0wa}, we display the corresponding constraints when CMB data are incorporated:
 \begin{equation}
    \begin{rcases}
    w_0 &= \text{ $-0.48 \pm 0.20$} \\
    w_a &= \text{$-1.57 \pm 0.58$} \\
\end{rcases}
\quad
\text{DESI1.5 + CMB (P+B)}.
\end{equation}

A comparable trend is observed: the addition of bispectrum information pulls the joint posterior toward \lcdm{}, consistent with findings from the DR1-only analysis \cite{novell-masot26}. However, the inclusion of DR2 BAO measurements substantially tightens the credible regions, and the resulting CMB+DESI combination exhibits a $2.8\sigma$\footnote{We approximate this frequentist figure from the difference in $\chi^2$ from the two models using $$\mathrm{CDF}_{\chi^2} (\Delta\chi^2_{\mathrm{MAP}} | 2\ \mathrm{dof}) = \frac{1}{\sqrt{2\pi}} \int_{-N}^{N} e^{-t^2/2} dt$$ as in other DESI papers.} tension with \lcdm{}—highlighting the enhanced constraining power of the expanded dataset while underscoring the persistent, albeit reduced, preference for dynamical dark energy in the joint CMB+ DESI analysis. We compute the Deviance Information Criterion (DIC)\footnote{We approximate the DIC value from the MCMC chains as $\mathrm{DIC} \approx \langle\chi^2\rangle + \frac{1}{2}\mathrm{var}(\chi^2)$, where ${\rm var}(\chi^2)$ denotes the variance of the $\chi^2$ values from the chain and $\langle\chi^2\rangle$ their mean. This is a fair approximation when the posteriors do not show significant projection effects and are in general roughly Gaussian.} change between models to be $\Delta \rm DIC \approx -7.8$, which indicates a strong preference (i.e. $-10 < \Delta \rm DIC < -5$) \cite{DIC} for \wowacdm{} even when somewhat accounting for the extra freedom penalty. 

In addition to these, we also report constraints including SNe information (which can also be seen in the bottom panels of \cref{fig:w0wa}).

For the SNe + DESI1.5 alone we find
  \begin{equation}
    \begin{rcases}
    w_0 &= \text{ $-0.882 \pm 0.065$} \\
w_a &= -0.34 \pm 0.30 \\
\end{rcases}
    \quad
    \text{DESI1.5 + BBN + $n_{s10}$ + DES-Dovekie (P+B)}.
    \end{equation}
These constraints are only $1.57\sigma$ away from \lcdm{}. This corresponds to a $\Delta{\rm DIC}\approx -0.7$ which indicates no preference for evolving DE from DESI and SNe. On the other hand, adding back the CMB information, pulls the contours away to $3.05\sigma$ or a $\Delta {\rm DIC} \approx -9.1$ which does suggest a strong preference for \wowacdm{} from this combination, implying that it is mostly the CMB information which drives the constraints away from \lcdm{}.
     \begin{equation}
    \begin{rcases}
    w_0 &= \text{$-0.817 \pm 0.057$} \\
w_a &= -0.68 \pm 0.21 \\
\end{rcases}
    \quad
    \text{DESI1.5 + CMB + DES-Dovekie (P+B)} 
    \end{equation}

A direct comparison with the Bayesian evidence analysis of \citep{ong2025} reveals differences in the inferred model preference. While our approximate DIC‑based results suggest a ``strong'' preference \cite{DIC} for \wowacdm{} when CMB is added, Ref.~\cite{ong2025} find that for a similar combination (BAO2 + CMB [Camspec] + DES‑Dovekie) the full evidence favors \lcdm{} ($\log B = -0.01\pm0.27$). Additionally, their results in the BAO2 + CMB combination shows the evidence for \wowacdm{} is weak at best regardless of the CMB likelihood chosen. They report a $\log B\approx 0.5$ (i.e. $1.5\sigma$ Bayesian significance) for Planck PR3 and $\log B\approx -0.6$ for Camspec (preference for \lcdm{}). This indicates that the approximate DIC penalty may be insufficient to fully account for the additional parameters of the \wowacdm{}, leading to an overstatement of the preference compared to the exact Bayesian evidence. Notably, the inclusion of the LRG bispectrum systematically shifts our dark energy constraints toward \lcdm{} (see \cref{tab:cosmo_params_w0wa} and \cref{fig:w0wa}), a trend consistent with independent supernova reanalyses \citep{DESreanalysis}. The bispectrum thus further mitigates this preference by breaking degeneracies that previously allowed the data to favor a more complex dark energy sector.

\subsubsection{Neutrino mass constraints}
\label{sec:nucdm_constraints}

\begin{figure}
    \centering
    \includegraphics[width=0.4\linewidth]{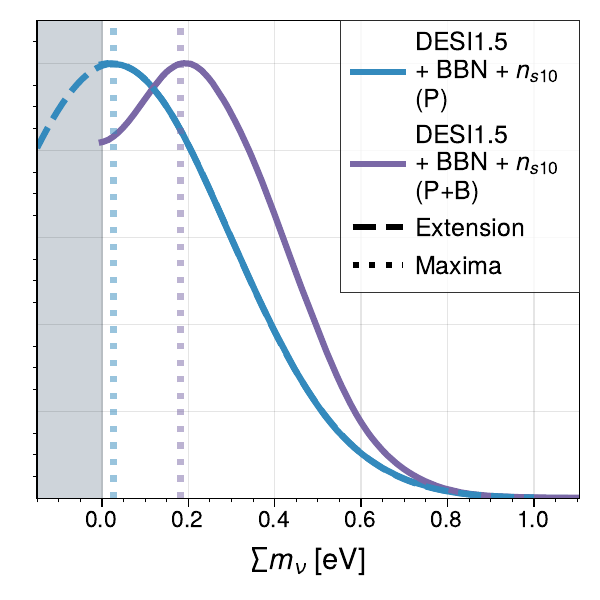}
    \includegraphics[width=0.4\linewidth]{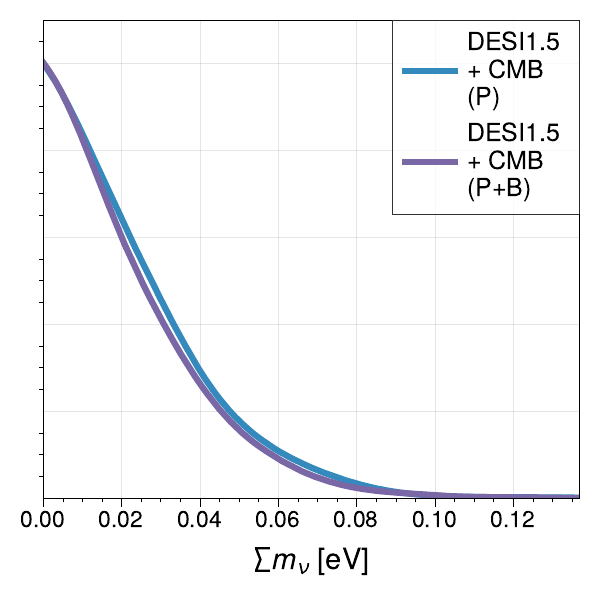}
    \includegraphics[width=0.7\linewidth]{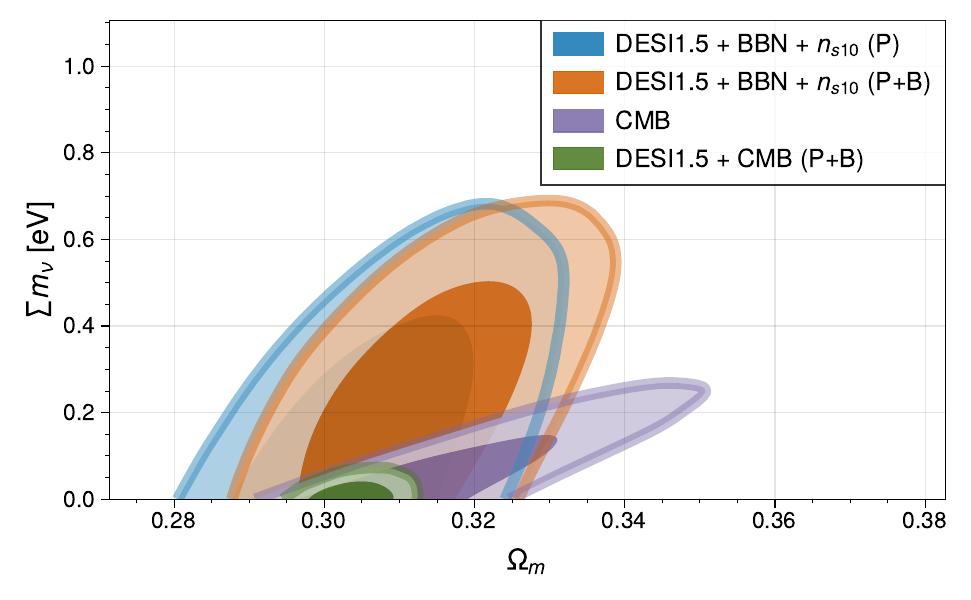}
    \caption{Neutrino mass constraints from the DESI1.5 dataset including the LRG bispectra compared to previous analyses without it. Top left: Constraints from DESI with minimal priors. Dashed lines show the extended Gaussian profile and dotted lines show the approximate MAP points. Top right: Constraints including CMB [Camspec]. Bottom: $\sum m_{\nu}$---$\Omega_m$ plane showing the degeneracy breaking between CMB and DESI1.5.}
    \label{fig:nucdm}
\end{figure}

The bispectrum is also sensitive to the late-time suppression from neutrino free-streaming and is thus an important observable for neutrino physics \cite{bispec_supp1,bispec_supp2,bispec_supp3,bispec_supp4,bispec_supp5}. We follow DESI's approach to neutrino constraints \cite{DESI2024.VII.KP7B, DESI2024.VI.KP7A} by allowing for three degenerate neutrino masses with a hard prior on $\sum m_\nu > 0$.
Our DESI‑only constraints (with BBN and $n_{s10}$ priors) illustrate this sensitivity. The power‑spectrum–only analysis (P) yields a marginalized mean of $\sum m_\nu = 0.23^{+0.059}_{-0.225}~\rm eV$, which is consistent with zero and has a long tail towards high masses. The addition of BAO2 to DR1 FS in \cite{desi1.5} moved the MAP towards slightly positive values, similarly, adding the bispectrum shifts the mean and 95\% CL to 
\begin{equation}
\label{eq:nucdm_constraints}
    \begin{rcases}
    \sum m_{\nu} = \text{ $0.26\pm0.18$}~\rm eV\quad\text{68\% CI}\\
    \sum m_{\nu} < \text{ $0.5926$}~\rm eV\quad\text{95\% CL}
\end{rcases}
    \quad
    \text{DESI1.5 + BBN + $n_{s10}$ (P+B), \lcdm{}} ,
    \end{equation}
further moving the posterior peak into the positive region. Consequently, the 95\% level also moves further away from zero. This can be observed in the top left panel of \cref{fig:nucdm}. Following the prescription of \cref{sec:results_combination}, the finite-mock uncertainty enters the quoted interval in two ways: the 95\% CL of \cref{eq:nucdm_constraints} includes the marginalized shift $\Delta_{95} = +0.0228~\rm eV$ (\cref{eq:delta95}) in order to account for the truncated posterior. The 68\% error includes the $0.18\sigma$ bootstrap scatter of the posterior mean added in quadrature, inflating it from $0.17~\rm eV$ to $0.18~\rm eV$. 

These neutrino mass constraints are still consistent with our previous power-spectrum-only analyses \cite{desi1.5} and with experiments such as KATRIN's constraint of \footnote{KATRIN cites upper bounds on a single neutrino mass of $m_\beta < 0.31~\rm eV\ \text{90\% CL}$ (Feldman-Cousins) which we very approximately translate into a sum of masses of $\sum m_\nu \lesssim 0.93~\rm eV$, very consistent with our constraint.} $\sum m_\nu \lesssim 0.93~\rm eV$ \cite{katrin}. Moreover, the recovered  peak in the positive mass range is in line with the lower mass bounds of $\sum m_\nu \gtrsim 0.058~\rm eV$ for normal ordering and $\sum m_\nu \gtrsim 0.098~\rm eV$ for inverted ordering provided in \cite{Nufit6} from neutrino oscillations.

The addition of CMB [Camspec] (top right in \cref{fig:nucdm}) dominates the neutrino mass constraints and drives the peak back to the prior boundary at $\sum m_\nu = 0$ regardless of the DESI1.5 data preference. The resulting constraints are tighter than without the bispectrum at 
\begin{equation}
    \sum m_{\nu} < \text{$0.0590$}~\rm eV\quad \text{95\% CL, DESI1.5  + CMB (P+B), \lcdm{}}
\end{equation}
This constraint is still driven by the $\sum m_\nu$---$\Omega_m$ degeneracy breaking between DESI and CMB data as illustrated in the bottom panel of \cref{fig:nucdm}. These tight constraints seem in tension with the lower bounds from neutrino oscillations \cite{Nufit6}, but are quite sensitive to the CMB likelihood used.  One might speculate that perhaps alternative CMB likelihoods, such as those which mitigate the lensing anomaly, may  alleviate this tension.  When combining with CMB, the finite-mock uncertainty is included in the quoted errors as elsewhere; it is largely subdominant with respect to the leading sources of information, i.e.\ the degeneracy differences and the CMB likelihood choice (e.g.\ the scatter on $\sum m_\nu$ is $0.09\sigma$ in \nulcdm{}+CMB), although it remains non-negligible for some parameters such as $\Omega_m$ and $h$ in that combination ($0.46\sigma$ and $0.44\sigma$). This upper limit is coincidentally the same as that found in \cite{Chudaykin2025b}, which used completely different analysis choices, demonstrating the robustness of the data.

\subsubsection{\nuwowacdm{} constraints}
\label{sec:nuw0wacdm_constraints}

\begin{figure}
    \centering
    \includegraphics[width=0.4\linewidth]{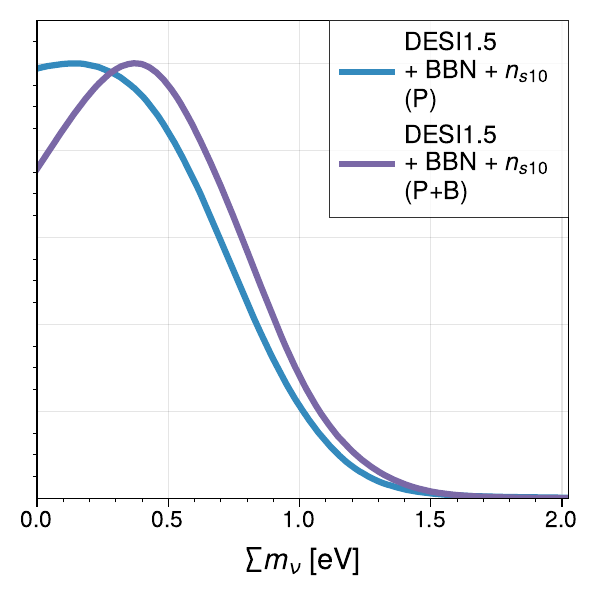}
    \includegraphics[width=0.4\linewidth]{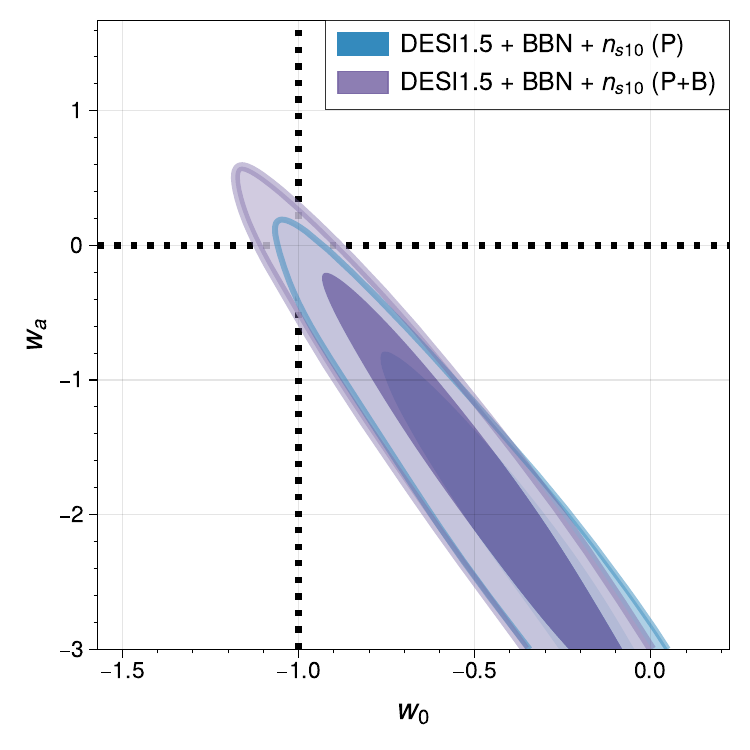}
    \caption{Cosmological constraints on a \nuwowacdm{} model with and without bispectra from DESI+BBN+$n_{s10}$. Left: Marginalized posteriors on the sum of neutrino masses, notice the shift towards positive values. Right: Constraints on DE parametrization showing the shift towards \lcdm{} induced by the addition of the bispectra.}
    \label{fig:nuwowa}
\end{figure}

When including evolving DE along with neutrinos we find the same trend in DE behavior, the bispectrum tends to shift contours towards the non-evolving limit of \lcdm{} (See \cref{fig:nuwowa}), albeit with increased uncertainties  due to the additional degree of freedom in the neutrino mass sum. The neutrino masses in turn also show a shift towards the positive mass end and we can therefore quote a constraint
\begin{equation}
\label{eq:nuw0wacdm_constraints}
    \begin{rcases}
    \sum m_{\nu} &= \text{$0.49 \pm 0.32$}~\mathrm{eV}\quad\text{68\% CI}\\
    \sum m_{\nu} &< \text{ $1.062$}~\mathrm{eV}\quad \text{95\% CL} \\
w_0 &= \text{$-0.56 \pm 0.26$} \\
w_a &= -1.54_{-1.21}^{+0.61} \\
\end{rcases}
    \quad
    \text{DESI1.5 + BBN + $n_{s10}$ (P+B)}.
    \end{equation}
Following the prescription of \cref{sec:results_combination}, the finite-mock uncertainty enters the quoted interval in two ways: the 95\% CL of \cref{eq:nuw0wacdm_constraints} includes the marginalized shift $\Delta_{95} = +0.0157~\rm eV$ (\cref{eq:delta95}) in order to account for the truncated posterior. The 68\% errors include the bootstrap scatter of the posterior mean added in quadrature; in relative units this corresponds to a $0.21\sigma$ inflation of the baseline error, which widens the quoted $\sum m_\nu$ and $w_0$ uncertainties by one unit in the last digit (to $\pm0.32$~eV and $\pm0.26$).

These are once again fully consistent with the expectations from neutrino oscillation experiments both in the upper and lower bounds.

\subsubsection{Constraints on spatial curvature}
\label{sec:olcdm_constraints}

Finally, we can explore the effect of the bispectrum information on the inferred constraints on the spatial curvature.

\begin{figure}
    \centering
    \includegraphics[width=0.49\linewidth]{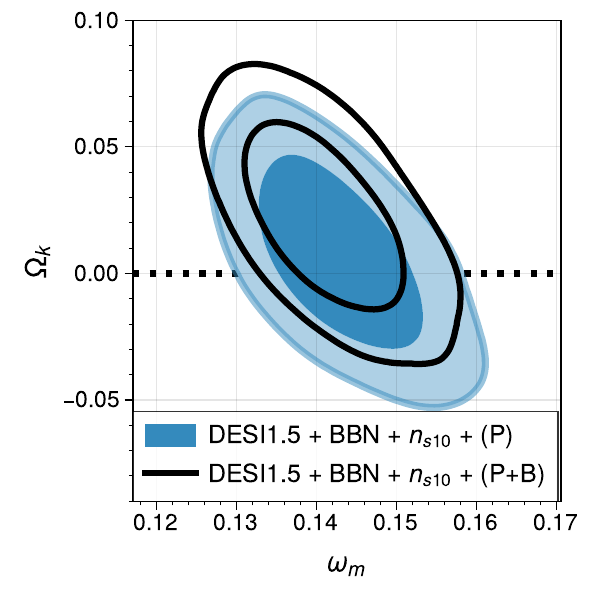}
    \includegraphics[width=0.49\linewidth]{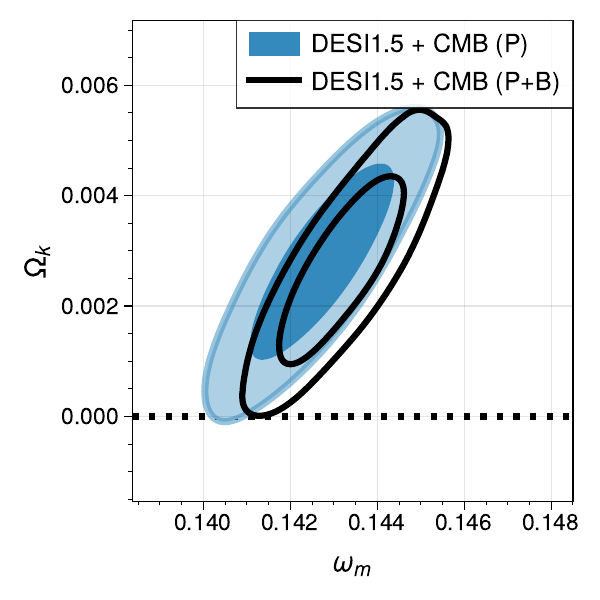}
    \caption{Constraints on the spatial curvature of the Universe from DESI1.5 with the bispectrum, compared to those without. Filled contours show results from \cite{desi1.5} using only the power spectrum, line contours show results from this work using the LRG bispectra.}
    \label{fig:omklcdm}
\end{figure}

\Cref{fig:omklcdm} shows the curvature constraints from DESI1.5 without CMB (left panel) and with CMB [Camspec] (right), plotted against the physical matter density. In our DESI-only constraints, the addition of the bispectrum shifts the central value towards positive $\Omega_k$ by $0.6\sigma$  while the marginalized uncertainty on $\Omega_k$ increases slightly from 0.024 to 0.025 after accounting for finite-mock effects ($0.20\sigma$).

\begin{table}
    \centering
    \resizebox{\columnwidth}{!}{\begin{tabular}{p{3cm}cccccc}
        \toprule
        Dataset & $\Omega_k$ & $h$ & $\Omega_m$ & $\sigma_8$ & $n_s$\\
\midrule 
\rowcolor[HTML]{EFEFEF}
DESI1.5 + BBN + $n_{s10}$ (P) & $0.007 \pm 0.024$  & $0.685 \pm 0.011$  & $0.3042 \pm 0.0090$  & $0.826 \pm 0.037$  & $0.970 \pm 0.039$ \\ 
\textbf{DESI1.5 + BBN + $n_{s10}$ (P+B)} & \boldmath$0.022 \pm 0.025$  & \boldmath$0.677 \pm 0.011$  & \boldmath$0.3078 \pm 0.0090$  & \boldmath$0.873 \pm 0.028$  & \boldmath$0.989 \pm 0.040$ \\ 
\rowcolor[HTML]{EFEFEF}
DESI1.5 + CMB (P) & $0.0028 \pm 0.0011$  & $0.6857 \pm 0.0030$  & $0.3035 \pm 0.0036$  & $0.8154_{-0.0055}^{+0.0051}$  & $0.9639 \pm 0.0037$ \\ 
\textbf{DESI1.5 + CMB (P+B)} & \boldmath$0.0027 \pm 0.0012$  & \boldmath$0.6828 \pm 0.0032$  & \boldmath$0.3072 \pm 0.0041$  & \boldmath$0.8177 \pm 0.0067$  & \boldmath$0.9630 \pm 0.0044$ \\ 
        \bottomrule
                        \end{tabular}}
\caption{Cosmological constraints for a model with curvature and different data combinations.}
    \label{tab:olcdm}
                    \end{table}

The shift toward positive $\Omega_k$ suppresses the growth of structure, which would reduce the predicted clustering amplitude. However, the bispectrum forces a higher observed amplitude, and the fit compensates by raising $\sigma_8$. The constraints on the full parameter space are shown in \cref{tab:olcdm}. Indeed, the constraint on $\sigma_8$ improves by  $24\%$ when the bispectrum is included. At the same time, $\Omega_m$ and $h$ shift by  $0.4\sigma$ and $0.7\sigma$, respectively, relative to the earlier DESI1.5 combination. Crucially, the physical matter density $\omega_m = h^2\Omega_m$ remains nearly unchanged (a $0.2\sigma$ shift) and shows a modest $4.3\%$ improvement in its uncertainty, a gain that is essentially unchanged once the finite-mock uncertainty on $h$ ($0.19\sigma$) is included. A modest improvement in the physical matter density is expected from the additional information that the bispectrum brings on the SF parameter $m$ \cite{Brieden21,Brieden2022,Brieden2023} whose bounds have been shown to be tightened when adding the bispectrum \cite{novell-masot25}.

In a joint analysis with the CMB, the  bispectrum  does not affect the curvature constraints, and the inference still shows a $2\sigma$ preference for positive $\Omega_k$. The constraint of the physical matter density  however is improved by 5.6\% at the statistical level, a gain that the finite-mock uncertainty more than offsets at the total-error level.

\subsubsection{Modified gravity}

\begin{figure}
    \centering
    \includegraphics[width=0.49\linewidth]{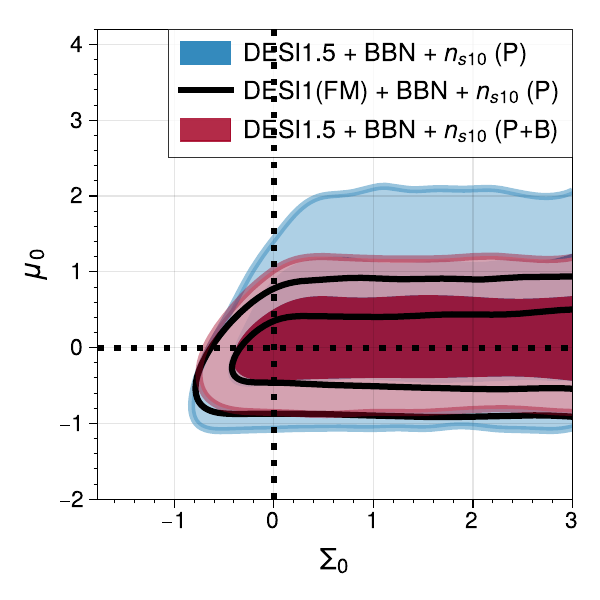}
    \includegraphics[width=0.49\linewidth]{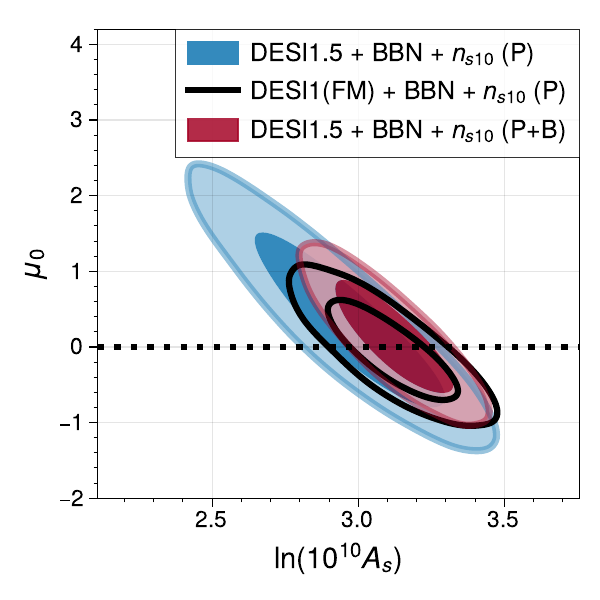}
    \includegraphics[width=0.49\linewidth]{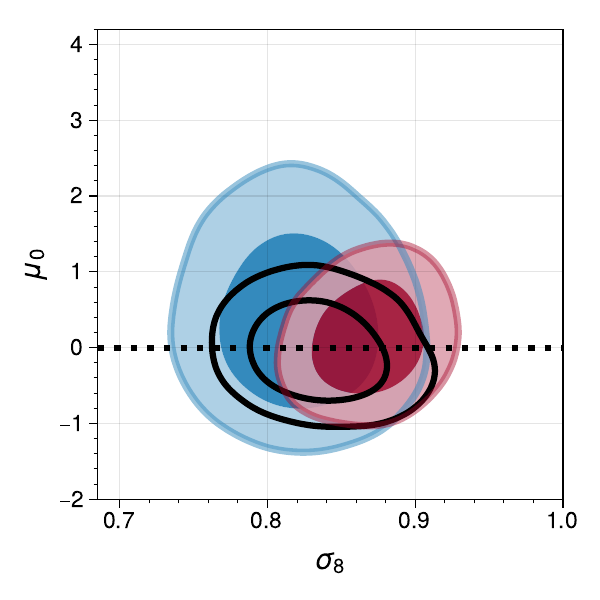}
    \caption{Modified gravity constraints from DESI1.5 and DESI1(FM) datasets for different parameter projections. Top left panel shows the MG parameter plane, showing an unconstrained $\Sigma_0$. The top right panel shows the degeneracy between the primordial amplitude $A_s$ and the $\mu_0$ parameter. The bottom panel shows the constraints on the late time amplitude parameter $\sigma_8$. Without the bispectrum, the SF approach measures only the degenerate combination $f\sigma_8$, so the amplitude parameters are less tightly constrained than in FM, where they are sampled directly and $f$ is derived from theory. Including the bispectrum breaks the growth-amplitude degeneracy, allowing SF to recover amplitude constraints that improve on those of FM.}
    \label{fig:mgcdm}
\end{figure}

\begin{table}
    \centering
    \resizebox{\columnwidth}{!}{\begin{tabular}{p{3cm}ccccccc}
        \toprule
        Dataset & $h$ & $\Omega_m$ & $\sigma_8$ & $n_s$ & $\ln(10^{10}A_s)$ & $\mu_0$ & $\Sigma_0$\\
\midrule 
\rowcolor[HTML]{EFEFEF}
\textbf{DESI1.5 + BBN + $n_{s10}$ (P)} & \boldmath$0.6871 \pm 0.0061$  & \boldmath$0.3033 \pm 0.0088$  & \boldmath$0.821 \pm 0.035$  & \boldmath$0.965 \pm 0.032$  & \boldmath$2.96 \pm 0.22$  & \boldmath$0.36 \pm 0.77$  & \bf{Unc.} \\ 
DESI1(FM) + BBN + $n_{s10}$ (P) & $0.6871 \pm 0.0077$  & $0.2946 \pm 0.0097$  & $0.835 \pm 0.031$  & $0.995 \pm 0.027$  & $3.11 \pm 0.15$  & $-0.04 \pm 0.44$  & Unc. \\ 
\rowcolor[HTML]{EFEFEF}
\textbf{DESI1.5 + BBN + $n_{s10}$ (P+B)} & \boldmath$0.6842 \pm 0.0060$  & \boldmath$0.3085 \pm 0.0086$  & \boldmath$0.869 \pm 0.027$  & \boldmath$0.967 \pm 0.031$  & \boldmath$3.12 \pm 0.14$  & \boldmath$0.12 \pm 0.51$  & \bf{Unc.} \\ 
        \bottomrule
                        \end{tabular}}
                        \caption{Constraints on the $\mu_0$--$\Sigma_0$ parametrization of modified gravity for DESI1.5 2- and 3-point based analyses. We quote the analog results from DESI1(FM) in \cite{DESI2024.VII.KP7B,KP7s1-MG}. The addition of the bispectrum improves the constraint on $\mu_0$ by around  34\% relative to the analogous SF-based constraint without it. When comparing to the P-only FM-based measurement, the constraint is  16\% worse. Boldface rows denote constraints new to this work.}
                        \label{tab:mgcdm}
                    \end{table}

The constraining power of the compressed SF approach in a modified gravity (MG) scenario is largely dependent on the parametrization. In this work we adopt the same $\mu_0$--$\Sigma_0$ parametrization (on top of the \lcdm{} parameters) used by earlier DESI studies on the matter using the FM approach \cite{DESI2024.VII.KP7B,KP7s1-MG}. This parametrization is a first-order approach to an effective modification of the gravity equations that relates the potentials $\Psi$ and $\Phi$ with the sources $\rho_i\Delta_i$ as
\begin{align}
    k^2\Psi &= -4\pi G a^2\mu(a,k)\sum\rho_i\Delta_i\\
    k^2(\Phi + \Psi) &= -8\pi G a^2\Sigma(a,k)\sum\rho_i\Delta_i.
\end{align}
The functions $\mu(a,k)$ and $\Sigma(a,k)$ are equal to 1 in general relativity (GR) but deviate in MG scenarios. In our approach we ignore the scale dependence and parametrize them both as
\begin{align}
    \label{eq:mudef}
    \mu(a,k) &= 1 + \mu_0\frac{\Omega_{\rm DE}}{\Omega_\Lambda}\\
    \Sigma(a,k) &= 1 + \Sigma_0\frac{\Omega_{\rm DE}}{\Omega_\Lambda},
\end{align}
and use $\mu_0$ and $\Sigma_0$ as our effective parameters. In GR these are both equal to 0. Given that we focus on DESI data only, the $\Sigma_0$ parameter is unconstrained as it does not enter the geodesic equations of matter particles, and therefore it does not affect the clustering of matter, DESI however should be able to constrain $\mu_0$ which modifies the growth of structure \cite{Pogosian2010}. Alternatively, MG analyses have traditionally parametrized deviations from GR in terms of the growth index $\gamma$, relating the growth rate and the density $f(z) = \Omega_m(z)^\gamma$. In GR it takes a value of $\gamma \approx 0.55$.

\Cref{fig:mgcdm} compares the 2D marginalized posteriors from DESI1.5 power‑spectrum–only (P), DESI1.5 with bispectrum (P+B), and the official DESI DR1 full‑modeling (FM) analysis \cite{DESI2024.VII.KP7B}. As expected, $\Sigma_0$ is completely unconstrained without lensing information (top‑left panel). Without the bispectrum, the SF approach only measures the degenerate combination $f\sigma_8$, hence it struggles to constrain $f$ independently from $\sigma_8$ as well as FM does, where amplitude parameters such as $\sigma_8$ (bottom panel) or $\log(10^{10}A_s)$ (top-right panel) are sampled directly and $f$ is derived from theory. Consequently, the amplitude parameters remain better constrained in the FM approach than in SF when the bispectrum is not added.

Adding the bispectrum in SF breaks the growth–amplitude degeneracy, allowing improved separate constraints on $f$ and $\sigma_8$, which improves the $\mu_0$ constraints by  $\sim 34\%$, when accounting for the uncertainty in our covariance matrix estimation through the addition of finite-mock uncertainties of  $0.21\sigma$ and $0.29\sigma$ to the $\log A_s$ and $\mu_0$ parameters (scatters of $0.43\sigma$ for $\sigma_8$ and $\leq0.22\sigma$ for the remaining parameters enter their total errors likewise)\footnote{In the IS run the finite-mock scatter is below $0.3\sigma$ for all parameters except $\sigma_8$ ($0.43\sigma$); since the quoted errors are the totals $\sigma_{\rm quad}$, $\sigma_8$ in MG ($\pm0.027$) is slightly looser than in \lcdm{} ($\pm0.026$).}, which have been computed using the importance sampling. Nevertheless, our $\mu_0$ constraint remains  $\sim16\%$ weaker than FM. This is likely because our bispectrum information is limited to the LRG samples, whereas FM uses the power‑spectrum shape of all tracers (including ELGs and QSOs) to constrain growth. In contrast, $\sigma_8$ is  $\sim13\%$ tighter in our P+B analysis than in FM, in line with the observations in other cosmological models. Likewise, $\log(10^{10}A_s)$ is $\sim7\%$ tighter than in FM. The DR2 BAO data (which only affect background parameters) do not contribute to these amplitude improvements. Though these constraints have been obtained through the Geo-FPT model, which has been calibrated to Newtonian gravity simulations, we expect changes in the SPT kernels due to modified gravity to be relatively small. Indeed, \cite{Bellini2015} has shown that for MG models with an expansion history consistent with GR, the bispectrum can be well modeled with GR kernels, and \cite{Gil2011} concluded that the gravity-model dependence is mostly encoded in the power spectrum. Thus, the main contribution of the bispectrum to MG constraints is degeneracy breaking in bias- and amplitude-related terms. The $\mu_0$ constraint should therefore be interpreted as conditional on the assumption that modified gravity affects the measured compressed P+B observables primarily through the linear growth quantities, and that modifications to the higher-order bispectrum kernels are subdominant.
Our full parameter constraints are summarized in \cref{tab:mgcdm}.

The constraints on $h$ and $\Omega_m$ are also improved due to the more constraining BAO2 information at a similar level as in the GR case. In terms of MG, we find
\begin{equation}
    \begin{rcases}
    \mu_0 &= \text{ $0.12 \pm 0.51$} \\
    \gamma &= 0.553 \pm 0.005
\end{rcases}
    \quad
    \text{DESI1.5 + BBN + $n_{s10}$ (P+B),\ MG},
    \end{equation}
where we have derived a growth index $\gamma$ value using \cite{Aparicio2017}
\begin{align}
    \gamma(\mu)&\approx\frac{1}{2} + \frac{0.161}{1.967 + \beta(\mu)}\\
    \beta(\mu)&\approx\frac{1}{4}\qty[\sqrt{1 + 24\mu} - 1].
\end{align}
where $\mu$ is given by \cref{eq:mudef} and would reduce to $\mu = 1 + \mu_0$ today. These constraints are consistent with GR ($\mu = 1$), where $\gamma = 0.554$ from this expression.
\section{Conclusions}
\label{sec:conclusions}
We have presented cosmological constraints from the combination of DESI Data Release 1 (DR1) full‑shape (FS) measurements, including the LRG bispectrum, and DESI Data Release 2 (DR2) baryon acoustic oscillation (BAO) measurements. The joint analysis properly accounts for the cross‑covariance between the two datasets using mock‑based estimates. The full‑shape information is compressed via the ShapeFit methodology, which mitigates prior volume effects that have hampered previous beyond‑\lcdm{} analyses. Crucially, by circumventing these projection effects, our analysis yields robust, DESI-only constraints in extended models. This allows us to rigorously assess DESI's independent constraining power as a standalone probe.

In the flat \lcdm{} model with minimal external priors (BBN and a loose $n_s$ prior), the present DESI1.5 (P+B) analysis favors a higher clustering amplitude than the previous power-spectrum-only DESI1.5 analysis of \cite{desi1.5}: $\sigma_8$ increases from $0.822\pm0.034$ to  $0.868\pm0.026$ while $S_8$ changes from $0.827\pm0.036$ to $0.880\pm0.028$. The marginalized uncertainties on these amplitude parameters improve by $24\%$ and $22\%$, respectively. Although the two DESI1.5 analyses are not identical apart from the addition of the bispectrum, the direction and improvement of the amplitude constraints are consistent with the controlled DR1 (P) versus (P+B) comparison of \cite{novell-masot26}, which demonstrated the ability of the bispectrum to break the galaxy-bias and matter-amplitude degeneracy. Background parameters ($h$, $\Omega_m$, $n_s$) show only marginal shifts, as their precision is dominated by the BAO data.

In the \wowacdm{} model with DESI-only data, the bispectrum shifts the dark energy
equation-of-state parameters toward the \lcdm{} point: $w_0$ moves from$-0.49\pm 0.25$ to $-0.64\pm 0.26$ and $w_a$ from $-1.52\pm 0.77$ to $-1.04\pm 0.82$. The resulting DESI1.5 constraints in the $w_0$–$w_a$ plane show no statistically significant preference for deviations from a cosmological constant. These findings are consistent with concurrent work with different modeling choices, like tree-level \cite{Chudaykin2025b} or one-loop bispectrum models \cite{Ivanov2026}, as well as differing prior choices, highlighting the data-driven nature of these observations (once prior volume effects are dealt with). 
When CMB data are included, the DESI+CMB (P+B) combination exhibits a $2.8\sigma$ (frequentist) preference for evolving dark energy, which decreases to $1.6\sigma$ for DESI+DES‑Dovekie and rises to $3.1\sigma$ for the full combination. We perform additional approximate DIC tests to take into account model complexity in our comparison and find that whenever CMB is included, the data prefers evolving DE, while combinations without CMB show no strong preference. This points towards the extra flexibility accommodating existing tensions between the datasets, such as the $\Omega_m$ tension between DESI and CMB. Furthermore, a comparison with full Bayesian evidence analyses on similar data combinations \cite{ong2025} indicates that these $\Delta\chi^2$-derived significances may overstate the preference because the DIC penalty approximates model complexity less severely than the exact Ockham factor. The addition of the bispectrum consistently shifts the posteriors toward \lcdm{}, weakening the evidence for time‑varying dark energy relative to power‑spectrum–only analyses.

The bispectrum provides enhanced sensitivity to the suppression of structure growth by massive neutrinos. In the DESI‑only analysis (BBN$+n_{s10}$), the power‑spectrum–only posterior for $\sum m_\nu$ is consistent with zero, whereas adding the bispectrum yields a mean of  $0.26\pm0.18~$eV and a $95\%$ upper limit of  $0.59~$eV, shifting the peak into the positive region and in agreement with  bounds set by oscillation experiments and with direct mass upper limits such as KATRIN. When CMB data are added, the constraint tightens dramatically ($\sum m_\nu<0.059~$eV at $95\%$ CL), driven by the strong complementarity between DESI and CMB. Even when the background is relaxed to include evolving dark energy, DESI alone constrains neutrino masses to $\sum m_\nu = 0.49\pm0.32~$eV, again consistent with oscillation bounds.

In terms of spatial curvature, in DESI‑only analyses the bispectrum shifts the mean of $\Omega_k$ from $0.007\pm0.024$ to  $0.022\pm0.025$, a $0.6\sigma$ shift, where the uncertainty on $\Omega_k$ now includes the finite-mock uncertainty ($0.20\sigma$). The physical matter density $\Omega_m h^2$ remains stable and slightly better constrained ($4.3\%$ improvement), as expected from the additional information the bispectrum brings to the ShapeFit parameter $m$. When CMB data are included, the constraints on curvature remain consistent with previous power-spectrum-only analyses and show a $\sim2\sigma$ preference for a positive $\Omega_k$.

In the modified gravity sector, the ShapeFit compression without the bispectrum only measures the degenerate combination $f\sigma_8$, hence it struggles to constrain $f$ independently from $\sigma_8$ as well as the FM approach does, where the amplitude is a sampled parameter and $f$ is derived from theory. Adding the LRG bispectrum (partially) breaks this degeneracy, improving the $\mu_0$ constraint by  $\sim34\%$ and yielding  $\mu_0 = 0.12\pm0.51$ ($68\%$~CI) from DESI‑only data which is fully consistent with general relativity (GR). Compared to full modeling (FM), our $\mu_0$ constraint remains  $\sim16\%$ weaker, likely because FM uses the power‑spectrum shape of all tracers (including ELGs and QSOs) to constrain growth (i.e. $f$), whereas our bispectrum information is limited to LRG samples. Nevertheless, our P+B analysis tightens $\sigma_8$ by  $\sim13\%$ and $\log(10^{10}A_s)$ by $\sim7\%$ relative to FM, in line with improvements seen in other cosmological models. Furthermore, we obtain an approximate $\gamma = 0.553 \pm 0.005$ which is also consistent with the GR value. The inclusion of the LRG bispectrum thus provides a robust handle on modified gravity from DESI alone, even with a compressed approach.

Our analysis builds on our previous work \cite{desi1.5,novell-masot25,novell-masot26} and demonstrates the power of combining the DR1 bispectrum with DR2 BAO while correctly accounting for their correlation. The ShapeFit compression allows robust Bayesian inference in extended models without the prior volume effects that affect full modeling approaches. The inclusion of the LRG bispectrum consistently shifts amplitude parameters upward, tightens their uncertainties, moves dark energy posteriors toward \lcdm{} and provides a handle on modified gravity from DESI alone with a compressed approach.

\section*{Data availability}
The data to reproduce the figures in this paper is available in the DESI \texttt{zenodo} repository \url{https://doi.org/10.5281/zenodo.20592803}. (Public after acceptance)

\section*{Acknowledgements}

DFS and HGM acknowledge support through the Consolidación Investigadora (CNS2023-144605) of the Spanish Ministry of Science and Innovation. HGM also acknowledges the support of the Ramón y Cajal (RYC-2021-034104).
Funding for this work was partially provided by the Spanish MINECO under project
PID2022-141125NB-I00 MCIN/AEI, and the “Center of Excellence Maria de Maeztu 2020-
2023” award to the ICCUB (CEX2019-000918-M funded by MCIN/AEI/10.13039/501100011033).

This material is based upon work supported by the U.S. Department of Energy (DOE), Office of Science, Office of High-Energy Physics, under Contract No. DE–AC02–05CH11231, and by the National Energy Research Scientific Computing Center, a DOE Office of Science User Facility under the same contract. Additional support for DESI was provided by the U.S. National Science Foundation (NSF), Division of Astronomical Sciences under Contract No. AST-0950945 to the NSF’s National Optical-Infrared Astronomy Research Laboratory; the Science and Technology Facilities Council of the United Kingdom; the Gordon and Betty Moore Foundation; the Heising-Simons Foundation; the French Alternative Energies and Atomic Energy Commission (CEA); the National Council of Humanities, Science and Technology of Mexico (CONAHCYT); the Ministry of Science, Innovation and Universities of Spain (MICIU/AEI/10.13039/501100011033), and by the DESI Member Institutions: \url{https://www.desi.lbl.gov/collaborating-institutions}. Any opinions, findings, and conclusions or recommendations expressed in this material are those of the author(s) and do not necessarily reflect the views of the U. S. National Science Foundation, the U. S. Department of Energy, or any of the listed funding agencies.

The authors are honored to be permitted to conduct scientific research on I'oligam Du'ag (Kitt Peak), a mountain with particular significance to the Tohono O’odham Nation.

\appendix

\section{Likelihood profiles}
\begin{figure}
    \centering
    \includegraphics[width=0.99\linewidth]{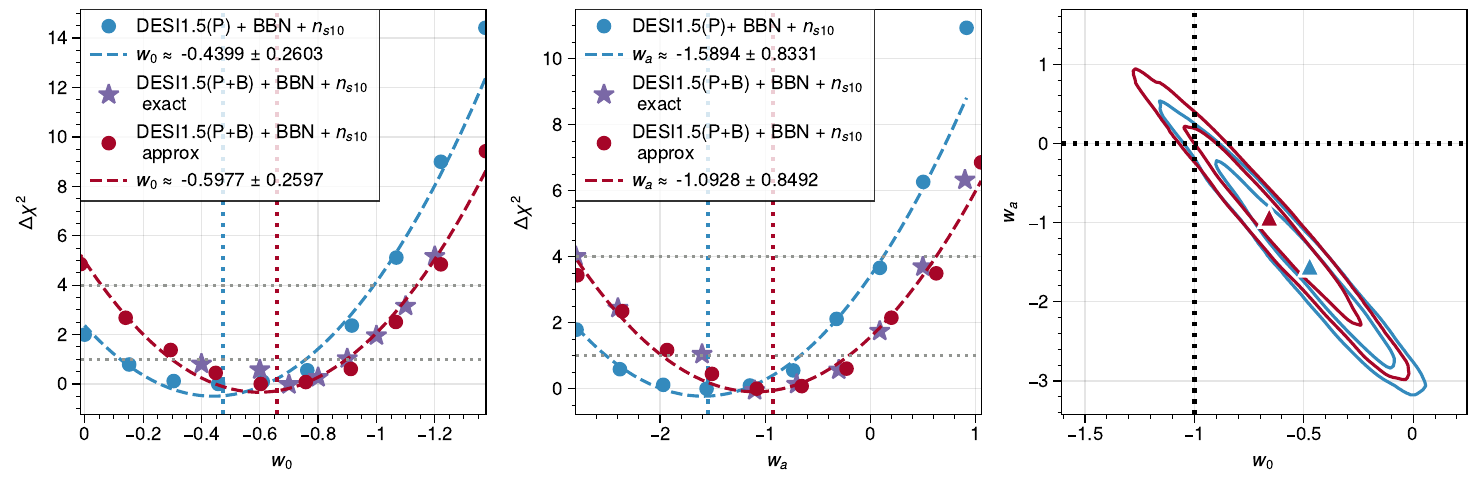}
    \includegraphics[width=0.99\linewidth]{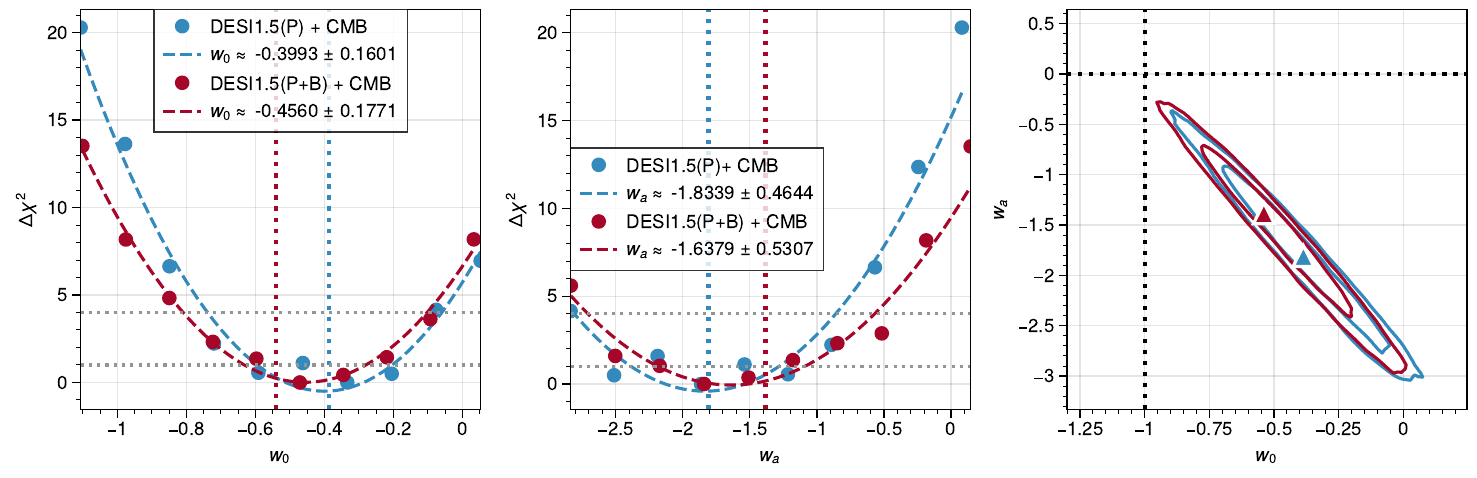}
    \caption{Binned $\chi^2$ profiles for DESI-only (top) and DESI+CMB (bottom) showing the absence of parameter projection effects when using SF with and without the bispectrum. The dashed lines in the 1D profiles correspond to a parabolic fit to the binned minimum $\chi^2$ (points) and the corresponding label shows the minimum from this parabolic fit. The stars \new{(DESI-only panels)} show exact minimizations of the posterior in good agreement with the points. The dotted vertical lines, on the other hand, show the global MAP values, which are in good agreement with the parabolae. The right panel shows a 2D KDE-estimate of the binned $\chi^2$ along with the global MAP values (triangles). Notice that these align with the Bayesian contours in \cref{fig:w0wa} confirming the absence of projection effects.}
    \label{fig:likeprof}
\end{figure}
Using the MCMC chains and their $\chi^2$ values we obtain binned estimates of the $\chi^2$ profiles by choosing the minimum value for each bin. This allows us to avoid expensive minimization runs and is a good estimate of the likelihood profile as its shape does not depend on the density of points (as the Bayesian marginals do) but on the likelihood values. \Cref{fig:likeprof} shows the estimated profiles (labeled ``approx'') along with a parabolic fit for the 1D profiles. From the parabola minima we estimate ``best-fit'' parameters that are reported in the legends. These are perfectly consistent with our Bayesian estimates reported in the main text and the MAP values from the full minimization shown as dotted vertical lines, showing that our approximation actually reproduces the likelihood profile. We stress that had the MCMC chains failed to sample the parameter region containing the true maximum, our approximate profile would have visibly deviated from the exact MAP, making the presence of projection effects immediately apparent. As an additional sanity check, we perform dedicated minimization runs on the DESI1.5+BBN+$n_{s10}$ likelihood, which are shown as stars and labeled ``exact''. The agreement between these exact minimizations and our chain-based approximation is remarkable, particularly for the least constraining likelihoods where projection effects are expected to be most severe. We thus conclude that our approximation is robust and skip computationally expensive minimizations for the 2D profiles and the more constraining likelihoods including CMB.

The 2D contours on the rightmost panels, correspond to Gaussian kernel density estimates of the binned $\chi^2$ surface, these also match the Bayesian contours in \cref{fig:w0wa} and encompass the MAP values (from 2D minimization) shown as triangles. We conclude that using SF produces constraints robust to prior projection effects.

\section{Uncertainty on cross-correlation matrix elements}
\label{app:cross-uncert}
\begin{figure}
    \centering
    \includegraphics[width=0.99\linewidth]{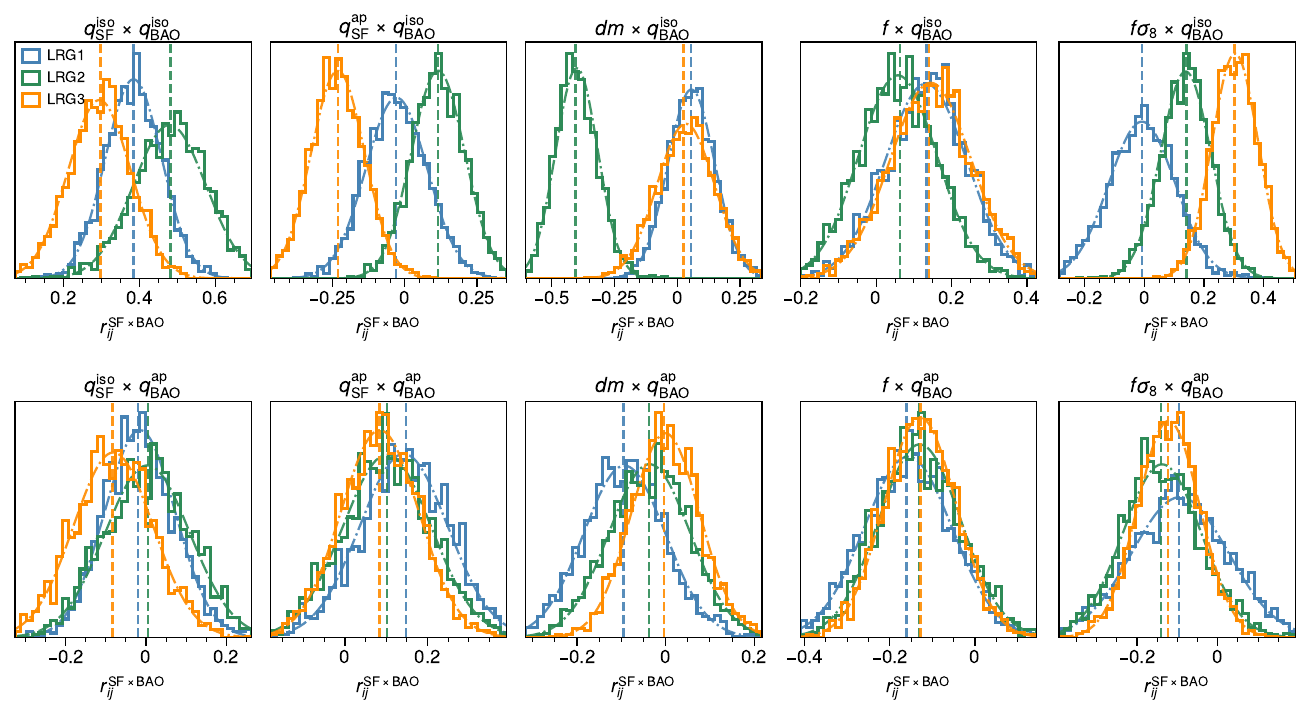}
    \caption{Bootstrap-estimated errors in the cross-correlation matrix terms. Histograms represent the distribution of cross correlation values and the vertical lines show the baseline values obtained from all mocks.}
    \label{fig:correlation_errors}
\end{figure}
In this appendix we estimate errors on the cross-correlation blocks computed from the mocks in compressed space, $\mathbf{r}_{\rm mock}^{\rm BAO\times SF}$. To estimate the errors we compute bootstrapped realizations of the correlation i.e., 500 to 2000 (depending on the dataset) draws with replacement from the 100 mocks, so that a random subset is repeated. 
 \Cref{fig:correlation_errors} presents these distributions for all the LRG bins, with the vertical lines marking the baseline values used in the paper. Their widths, $\sim0.08$--$0.12$, agree with the expected sampling error $(1-r^2)/\sqrt{N_{\rm Mocks}-1}\simeq0.08$--$0.10$. We evaluate the impact of these variations in cosmological constraints in the next section. \Cref{tab:cross_corr} shows the baseline numerical values for the cross correlation along with their lower and upper 1$\sigma$ values computed as the 16th and 84th percentiles.

\begin{table}[htbp]
  \centering
  \small
  \begin{tabular}{lccccc}
    \hline\hline
    \multicolumn{6}{c}{\textbf{Cross-correlation coefficients} $r_{ij} \pm 1\sigma$} \\
    \hline
    \multicolumn{6}{c}{\textit{LRG1}} \\
    & $q_{\rm SF}^{\rm iso}$ & $q_{\rm SF}^{\rm ap}$ & $dm$ & $f$ & $f\sigma_8$ \\
    \hline
    $q_{\rm BAO}^{\rm iso}$ & $0.38_{-0.07}^{+0.08}$ & $-0.03_{-0.11}^{+0.11}$ & $0.06_{-0.09}^{+0.09}$ & $0.13_{-0.11}^{+0.11}$ & $-0.01_{-0.11}^{+0.11}$ \\
    $q_{\rm BAO}^{\rm ap}$ & $-0.02_{-0.08}^{+0.10}$ & $0.15_{-0.11}^{+0.10}$ & $-0.10_{-0.09}^{+0.09}$ & $-0.16_{-0.10}^{+0.11}$ & $-0.10_{-0.12}^{+0.13}$ \\
    \hline
    \multicolumn{6}{c}{\textit{LRG2}} \\
    & $q_{\rm SF}^{\rm iso}$ & $q_{\rm SF}^{\rm ap}$ & $dm$ & $f$ & $f\sigma_8$ \\
    \hline
    $q_{\rm BAO}^{\rm iso}$ & $0.48_{-0.10}^{+0.09}$ & $0.11_{-0.09}^{+0.10}$ & $-0.40_{-0.08}^{+0.09}$ & $0.06_{-0.11}^{+0.10}$ & $0.14_{-0.09}^{+0.08}$ \\
    $q_{\rm BAO}^{\rm ap}$ & $0.01_{-0.11}^{+0.11}$ & $0.10_{-0.11}^{+0.11}$ & $-0.04_{-0.09}^{+0.09}$ & $-0.13_{-0.10}^{+0.10}$ & $-0.14_{-0.09}^{+0.10}$ \\
    \hline
    \multicolumn{6}{c}{\textit{LRG3}} \\
    & $q_{\rm SF}^{\rm iso}$ & $q_{\rm SF}^{\rm ap}$ & $dm$ & $f$ & $f\sigma_8$ \\
    \hline
    $q_{\rm BAO}^{\rm iso}$ & $0.30_{-0.08}^{+0.08}$ & $-0.23_{-0.09}^{+0.09}$ & $0.02_{-0.11}^{+0.12}$ & $0.14_{-0.11}^{+0.11}$ & $0.30_{-0.08}^{+0.08}$ \\
    $q_{\rm BAO}^{\rm ap}$ & $-0.08_{-0.10}^{+0.10}$ & $0.08_{-0.10}^{+0.09}$ & $-0.00_{-0.08}^{+0.08}$ & $-0.13_{-0.09}^{+0.08}$ & $-0.12_{-0.08}^{+0.08}$ \\
    \hline
    \hline\hline
  \end{tabular}
  \caption{Fiducial SF$\times$BAO cross-correlation coefficients $r_{ij}$ and their two-sided 1$\sigma$ uncertainties from the mock bootstrap. Upper errors correspond to the 84th percentile, lower errors to the 16th.}
  \label{tab:cross_corr}
\end{table}

\section{Robustness of cosmology constraints to changes in cross-covariance matrices}
\label{app:robustness_cosmo}
\begin{figure}
    \centering
    \includegraphics[width=0.99\linewidth]{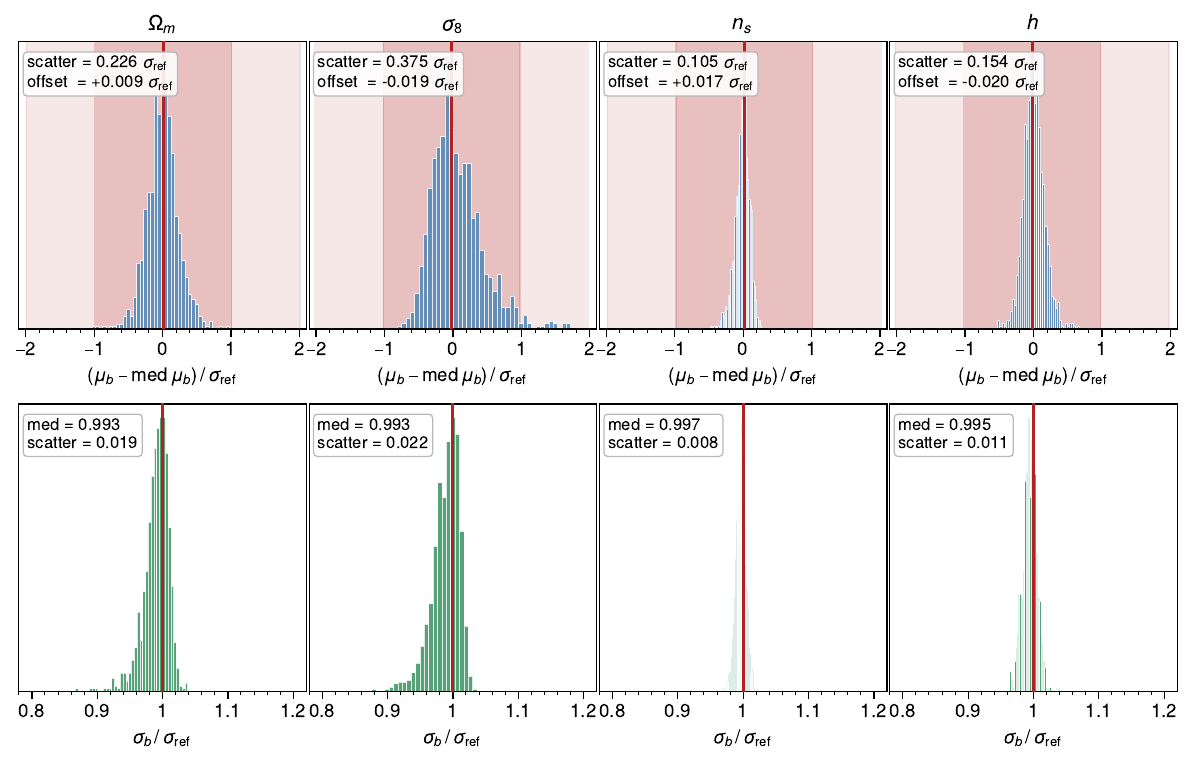}
    \caption{Bootstrap-estimate robustness of the cosmological parameters to changes in the SF$\times$BAO cross-correlation coefficients. Top row shows the distributions of parameter shift estimates $\mu_b - \langle\mu_b\rangle$ from 1000 different importance-sampled chains in units of the baseline statistical error $\sigma_{\rm ref}$. The bottom row shows the ratio of estimated parameter errors $\sigma_b$ relative to the baseline estimate. The red vertical lines show the baseline statistics from \cref{tab:lcdm_tab}.}
    \label{fig:param_robustness}
\end{figure}

We further evaluate the effect of the uncertainties in the LRG covariances by importance sampling our DESI1.5+$n_{s10}$+BBN chain by weighting each sample with 
\begin{align}
    \Delta\log \mathcal{L} = -\frac{1}{2}\sum_{i \in \rm LRG}\qty[\hat{\Theta}_{i,\rm comp}^{*,\,\rm data} - \Theta_{\rm comp}^{\rm model}]^T\Delta\qty(\mathbf{\hat{C}}^{\rm BAO + SF}_{\rm comb})_i^{-1}\qty[\hat{\Theta}_{\rm comp}^{*,\,\rm data} - \Theta_{i,\rm comp}^{\rm model}],
\end{align}
where
\begin{equation}
    \Delta\qty(\mathbf{\hat{C}}^{\rm BAO + SF}_{\rm comb})_{(i,b)}^{-1} = \qty(\mathbf{\hat{C}}^{\rm BAO + SF}_{\rm comb,boot})_{(i,b)}^{-1} - \qty(\mathbf{\hat{C}}^{\rm BAO + SF}_{\rm comb})_i^{-1},
\end{equation}
and the $b$ index labels the $b$-th bootstrap realization. Note that the change in the Gaussian normalization, $-\tfrac{1}{2}\Delta\ln\det\mathbf{\hat{C}}^{\rm BAO+SF}_{\rm comb}$, is independent of the cosmological parameters and therefore cancels when the importance weights are self-normalized. With this procedure we generate importance-sampled chains and measure the cosmological parameters as the means of the posterior distributions ($\mu_b$) as well as their estimated 1$\sigma$ errors, $\sigma_b$. We have ensured the median effective sample size of the bootstrap chains remains over 5\% of the total sample size. \Cref{fig:param_robustness} shows the distributions of these statistics for \lcdm{} (as a reference). The top row shows the standardized distribution of parameter estimates in units of $\sigma_{\rm ref}$ i.e. the error bars obtained from the baseline analysis.  For the observed data, this is the analogue of the excess scatter of best-fit parameters identified by \cite{DodelsonSchneider2013}, evaluated for our hybrid covariance matrix. The scatter of the cosmological estimates from the bootstrapped chains is comfortably below the reference error for most of the parameters. Nonetheless, $\sigma_8$ shows a scatter of $\sim 0.4\sigma_{\rm ref}$, this means that it would be reasonable to find changes of that much due only to the uncertainty in the cross correlation values. 

The bottom row of \cref{fig:param_robustness} shows the ratios of the estimated parameter errors relative to the baseline. These are highly concentrated around 1: the posterior width itself is relatively insensitive to the cross-covariance uncertainty, which therefore enters mainly through the scatter of the mean shown in the top row.

\section{Author Affiliations}
\label{sec:affiliations}

\noindent \hangindent=.5cm $^{a}${Institut de Ci\`encies del Cosmos (ICCUB), Universitat de Barcelona (UB), c. Mart\'i i Franqu\`es, 1, 08028 Barcelona, Spain.}

\noindent \hangindent=.5cm $^{b}${Departament de F\'{\i}sica Qu\`{a}ntica i Astrof\'{\i}sica, Universitat de Barcelona, Mart\'{\i} i Franqu\`{e}s 1, E08028 Barcelona, Spain}

\noindent \hangindent=.5cm $^{c}${Institut d'Estudis Espacials de Catalunya (IEEC), c/ Esteve Terradas 1, Edifici RDIT, Campus PMT-UPC, 08860 Castelldefels, Spain}

\noindent \hangindent=.5cm $^{d}${Instituci\'{o} Catalana de Recerca i Estudis Avan\c{c}ats, Passeig de Llu\'{\i}s Companys, 23, 08010 Barcelona, Spain}

\noindent \hangindent=.5cm $^{e}${Lawrence Berkeley National Laboratory, 1 Cyclotron Road, Berkeley, CA 94720, USA}

\noindent \hangindent=.5cm $^{f}${Department of Physics, Boston University, 590 Commonwealth Avenue, Boston, MA 02215 USA}

\noindent \hangindent=.5cm $^{g}${Dipartimento di Fisica ``Aldo Pontremoli'', Universit\`a degli Studi di Milano, Via Celoria 16, I-20133 Milano, Italy}

\noindent \hangindent=.5cm $^{h}${INAF-Osservatorio Astronomico di Brera, Via Brera 28, 20122 Milano, Italy}

\noindent \hangindent=.5cm $^{i}${Department of Physics \& Astronomy, University College London, Gower Street, London, WC1E 6BT, UK}

\noindent \hangindent=.5cm $^{j}${Institute of Space Sciences, ICE-CSIC, Campus UAB, Carrer de Can Magrans s/n, 08913 Bellaterra, Barcelona, Spain}

\noindent \hangindent=.5cm $^{k}${Institute for Computational Cosmology, Department of Physics, Durham University, South Road, Durham DH1 3LE, UK}

\noindent \hangindent=.5cm $^{l}${Instituto de F\'{\i}sica, Universidad Nacional Aut\'{o}noma de M\'{e}xico,  Circuito de la Investigaci\'{o}n Cient\'{\i}fica, Ciudad Universitaria, Cd. de M\'{e}xico  C.~P.~04510,  M\'{e}xico}

\noindent \hangindent=.5cm $^{m}${NSF NOIRLab, 950 N. Cherry Ave., Tucson, AZ 85719, USA}

\noindent \hangindent=.5cm $^{n}${Department of Astronomy \& Astrophysics, University of Toronto, Toronto, ON M5S 3H4, Canada}

\noindent \hangindent=.5cm $^{o}${Department of Physics \& Astronomy and Pittsburgh Particle Physics, Astrophysics, and Cosmology Center (PITT PACC), University of Pittsburgh, 3941 O'Hara Street, Pittsburgh, PA 15260, USA}

\noindent \hangindent=.5cm $^{p}${University of California, Berkeley, 110 Sproul Hall \#5800 Berkeley, CA 94720, USA}

\noindent \hangindent=.5cm $^{q}${Institut de F\'{i}sica d’Altes Energies (IFAE), The Barcelona Institute of Science and Technology, Edifici Cn, Campus UAB, 08193, Bellaterra (Barcelona), Spain}

\noindent \hangindent=.5cm $^{r}${Departamento de F\'isica, Universidad de los Andes, Cra. 1 No. 18A-10, Edificio Ip, CP 111711, Bogot\'a, Colombia}

\noindent \hangindent=.5cm $^{s}${Observatorio Astron\'omico, Universidad de los Andes, Cra. 1 No. 18A-10, Edificio H, CP 111711 Bogot\'a, Colombia}

\noindent \hangindent=.5cm $^{t}${University of Virginia, Department of Astronomy, Charlottesville, VA 22904, USA}

\noindent \hangindent=.5cm $^{u}${Fermi National Accelerator Laboratory, PO Box 500, Batavia, IL 60510, USA}

\noindent \hangindent=.5cm $^{v}${Department of Astronomy, University of Texas at Austin, 2515 Speedway, TX 78712, USA}

\noindent \hangindent=.5cm $^{w}${Institut d'Astrophysique de Paris. 98 bis boulevard Arago. 75014 Paris, France}

\noindent \hangindent=.5cm $^{x}${IRFU, CEA, Universit\'{e} Paris-Saclay, F-91191 Gif-sur-Yvette, France}

\noindent \hangindent=.5cm $^{y}${Center for Cosmology and AstroParticle Physics, The Ohio State University, 191 West Woodruff Avenue, Columbus, OH 43210, USA}

\noindent \hangindent=.5cm $^{z}${Department of Physics, The Ohio State University, 191 West Woodruff Avenue, Columbus, OH 43210, USA}

\noindent \hangindent=.5cm $^{aa}${The Ohio State University, Columbus, 43210 OH, USA}

\noindent \hangindent=.5cm $^{ab}${Department of Physics, University of Michigan, 450 Church Street, Ann Arbor, MI 48109, USA}

\noindent \hangindent=.5cm $^{ac}${University of Michigan, 500 S. State Street, Ann Arbor, MI 48109, USA}

\noindent \hangindent=.5cm $^{ad}${Department of Physics, The University of Texas at Dallas, 800 W. Campbell Rd., Richardson, TX 75080, USA}

\noindent \hangindent=.5cm $^{ae}${Department of Physics and Astronomy, University of California, Irvine, 92697, USA}

\noindent \hangindent=.5cm $^{af}${Sorbonne Universit\'{e}, CNRS/IN2P3, Laboratoire de Physique Nucl\'{e}aire et de Hautes Energies (LPNHE), FR-75005 Paris, France}

\noindent \hangindent=.5cm $^{ag}${Departament de F\'{i}sica, Serra H\'{u}nter, Universitat Aut\`{o}noma de Barcelona, 08193 Bellaterra (Barcelona), Spain}

\noindent \hangindent=.5cm $^{ah}${Department of Physics and Astronomy, Siena University, 515 Loudon Road, Loudonville, NY 12211, USA}

\noindent \hangindent=.5cm $^{ai}${Institute of Cosmology and Gravitation, University of Portsmouth, Dennis Sciama Building, Portsmouth, PO1 3FX, UK}

\noindent \hangindent=.5cm $^{aj}${Departamento de F\'{\i}sica, DCI-Campus Le\'{o}n, Universidad de Guanajuato, Loma del Bosque 103, Le\'{o}n, Guanajuato C.~P.~37150, M\'{e}xico}

\noindent \hangindent=.5cm $^{ak}${Instituto Avanzado de Cosmolog\'{\i}a A.~C., San Marcos 11 - Atenas 202. Magdalena Contreras. Ciudad de M\'{e}xico C.~P.~10720, M\'{e}xico}

\noindent \hangindent=.5cm $^{al}${Department of Physics and Astronomy, University of Waterloo, 200 University Ave W, Waterloo, ON N2L 3G1, Canada}

\noindent \hangindent=.5cm $^{am}${Perimeter Institute for Theoretical Physics, 31 Caroline St. North, Waterloo, ON N2L 2Y5, Canada}

\noindent \hangindent=.5cm $^{an}${Waterloo Centre for Astrophysics, University of Waterloo, 200 University Ave W, Waterloo, ON N2L 3G1, Canada}

\noindent \hangindent=.5cm $^{ao}${Instituto de Astrof\'{i}sica de Andaluc\'{i}a (CSIC), Glorieta de la Astronom\'{i}a, s/n, E-18008 Granada, Spain}

\noindent \hangindent=.5cm $^{ap}${Departament de F\'isica, EEBE, Universitat Polit\`ecnica de Catalunya, c/Eduard Maristany 10, 08930 Barcelona, Spain}

\noindent \hangindent=.5cm $^{aq}${Department of Physics and Astronomy, Sejong University, 209 Neungdong-ro, Gwangjin-gu, Seoul 05006, Republic of Korea}

\noindent \hangindent=.5cm $^{ar}${Abastumani Astrophysical Observatory, Tbilisi, GE-0179, Georgia}

\noindent \hangindent=.5cm $^{as}${Department of Physics, Kansas State University, 116 Cardwell Hall, Manhattan, KS 66506, USA}

\noindent \hangindent=.5cm $^{at}${Faculty of Natural Sciences and Medicine, Ilia State University, 0194 Tbilisi, Georgia}

\noindent \hangindent=.5cm $^{au}${CIEMAT, Avenida Complutense 40, E-28040 Madrid, Spain}

\noindent \hangindent=.5cm $^{av}${Institut Qu\'imic de Sarri\`a-Universitat Ramon Llull (IQS-URL), Via Augusta 390, 08017, Barcelona, Spain}

\bibliographystyle{JHEP} 
\bibliography{refs,Misha_DESI_supporting_papers2025-05-11} 
\end{document}